\documentclass{PoS}

\usepackage{graphics}
\usepackage{multirow}
\usepackage{xspace}
\usepackage{amsmath}
\usepackage{amssymb}
\usepackage[utf8]{inputenc}
\usepackage[T1]{fontenc}
\usepackage[switch]{lineno}
\usepackage{enumerate}
\RequirePackage{graphicx}
\usepackage{color}
\usepackage{xspace}

\newcommand{\VTPCOne}{{VTPC-1}\xspace}
\newcommand{\VTPCTwo}{{VTPC-2}\xspace}
\newcommand{\GAPTPC}{{GAP-TPC}\xspace}

\newcommand{\MTPCL}{{MTPC-L}\xspace}
\newcommand{\MTPCR}{{MTPC-R}\xspace}

\newcommand{\dedx}{\ensuremath{{\rm d}E\!/\!{\rm d}x}\xspace}

\newcommand{\GeV}{\ensuremath{\mbox{Ge\kern-0.1em V}}\xspace}
\newcommand{\GeVc}{\ensuremath{\mbox{Ge\kern-0.1em V}\!/\!c}\xspace}
\newcommand{\GeVcc}{\ensuremath{\mbox{Ge\kern-0.1em V}\!/\!c^2}\xspace}
\newcommand{\AGeV}{\ensuremath{A\,\mbox{Ge\kern-0.1em V}}\xspace}
\newcommand{\AGeVc}{\ensuremath{A\,\mbox{Ge\kern-0.1em V}\!/\!c}\xspace}

\newcommand{\EposLong}{{\scshape Epos1.99}\xspace}

\newcommand{\NASixtyOne}{\mbox{NA61\slash SHINE}\xspace}%this seems to work properly to me. aa

\title{Recent results from NA61/SHINE on spectra and correlations in p+p and Be+Be interactions at the CERN SPS}

\ShortTitle{Recent results from NA61/SHINE on spectra and correlations in p+p and Be+Be interactions at the CERN SPS}

\author{\speaker{Andrzej Wilczek} \textbf{\textit{(for the NA61/SHINE Collaboration)}}\\ 
        University of Silesia in Katowice, Poland\\
        E-mail: \email{awilczek@us.edu.pl}}

\abstract{The problem of pinning down the critical point of strongly interacting
matter still puzzles the community. One of the answers suspected
to emerge in the near future will surely come from NA61/SHINE -
a fixed-target experiment aiming to discover the critical point as well
as to study the properties of the onset of deconfinement.

This goal will be pursued by obtaining precise data on hadron
production in proton-proton, proton-nucleus and nucleus-nucleus
interactions in a wide range of system size and collision energy.

This contribution presents new results on inclusive spectra of identified hadrons
and on fluctuations in inelastic p+p and Be+Be interactions
at the SPS energies.
These are compared with the world data, in particular
with the corresponding measurements of NA49 for central Pb+Pb collisions
as well as with some model predictions.}

\FullConference{ 7th International Conference on Physics and Astrophysics of Quark Gluon Plasma\\
 1-5 February , 2015 \\
 Kolkata, India}

\begin{document}
\sloppy
\section{The NA61/SHINE facility}

\begin{figure*}
\begin{center}
\resizebox{0.85\textwidth}{!}{
  \includegraphics{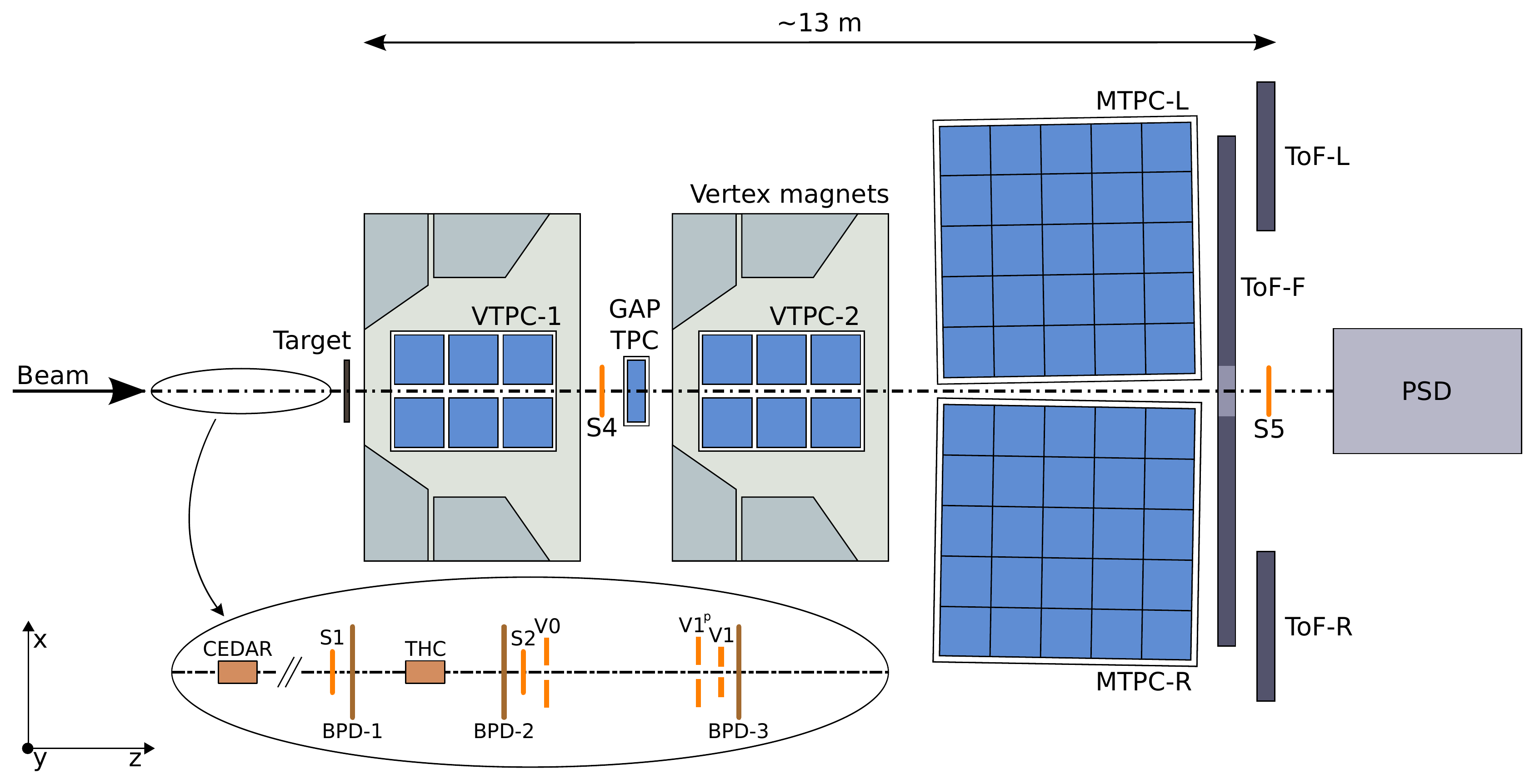}
}
\end{center}
\caption{Schematic layout of the NA61/SHINE experiment at the CERN SPS
(horizontal cut in the beam plane, not to scale).
The chosen
right-handed coordinate system is shown on the plot.
The incoming beam direction is along the z axis.
The magnetic field bends charged particle trajectories in the x-z
(horizontal) plane.
The drift direction in the TPCs is along the y (vertical) axis~\cite{Facility}.}
\label{fig:na61} % Give a unique lebel
\end{figure*}

The \NASixtyOne experiment~\cite{Facility} uses a large 
acceptance hadron spectrometer located
in the H2 beam-line at the CERN SPS accelerator complex. The layout of the
experiment is schematically shown in Fig.~\ref{fig:na61}. 
The main detector system 
is a set of large volume Time Projection Chambers (TPCs).
Two of them (\mbox{\VTPCOne} and \mbox{\VTPCTwo}) are placed 
inside super-conducting magnets
(\mbox{VTX-1} and \mbox{VTX-2})
with a combined bending power of 9~Tm. The standard current setting for
data taking
at 158~\GeVc corresponds to full field, 1.5~T, in the first and reduced field,
1.1~T, in the second magnet. For lower beam momenta the field is scaled by $p_{beam}/158$, where $p_{beam}$ is 
the beam momentum expressed in \AGeVc.

Two large TPCs (\MTPCL and \MTPCR) are positioned downstream of the magnets, symmetrically to the undeflected beam.
A fifth small TPC (\GAPTPC) is placed between \mbox{\VTPCOne} 
and \mbox{\VTPCTwo} directly
on the beam line and
covers the gap between the sensitive volumes of the other TPCs. The \NASixtyOne
TPC system allows
precise measurement of the particle momenta $p$ with a resolution
of $\sigma(p)/p^2\approx (0.3 - 7)\!\times\!10^4\;\rm{(GeV/c)}^{-1}$ 
at the full magnetic field used for data taking at 158~\GeVc and provides
particle identification
via the measurement of the specific energy loss, \dedx, with relative
resolution of about 4.5\%.

A set of scintillation and Cherenkov counters, as well as beam position
detectors (BPDs) upstream of the main detection system provide the timing
reference, as well as identification and position measurements of the incoming beam
particles. 

Secondary hadron beams of momentum ranging from 20 to 158 \GeVc are produced by 
400~\GeVc primary protons impinging on a 10~cm long beryllium target. Hadrons produced at the
target are transported downstream to the \NASixtyOne experiment along the
H2 beamline, in which collimation and momentum selection occur. Protons
in the secondary hadron beam are identified by a differential Cherenkov
counter (CEDAR).   
%Two scintillation counters, S1 and S2,
%together with the three veto counters V0, V1 and V1$^p$ were used to select
%beam particles. Thus, the selected beam protons were required to satisfy the coincidence
%$S1\cdot S2\cdot\overline{V0}\cdot\overline{V1}\cdot\overline{V1^p}\cdot
%CEDAR$. Trajectories of individual beam particles were measured in the
%telescope of beam position detectors placed along the beam line (\BPDAll in
%Fig.~\ref{fig:na61}). These are multiwire proportional chambers
%with two orthogonal sense wire planes and cathode strip readout, allowing
%to determine the transverse coordinates of the individual beam particle at
%the target position with a resolution of about 100~$\mu$m.
For data taking on p+p interactions a liquid hydrogen target (LHT) 
of 20.29~cm length (2.8\% interaction length) and 3~cm diameter 
was placed 88.4~cm upstream of \VTPCOne. 
%The boiling rate of hydrogen was not monitored during
%the data acquisition, thus the density is known only approximately. It has
%however no impact on the results presented in this paper as they are determined
%from particle yields per selected event and thus are independent of the target
%density. 
Inelastic p+p interactions in the LHT are
selected by requiring an anti-coincidence of an identified incoming beam proton with a
small scintillation counter of 2~cm diameter (S4) placed on the beam trajectory
between the two spectrometer magnets. 

Be beams are obtained by fragmentation of primary Pb ions from the SPS in a Be target of
18 cm length. The magnets and collimators of the H2 beam line are set to select a clean 
$^7$Be beam from the fragmentation products. This beam is impinging on a $^9$Be plate
target of 1.2 cm thickness.
Trigger and centrality selection in $^7$Be+$^9$Be interactions were performed using a modular zero-degree calorimeter
(Particle Spectator Detector - PSD). The selection is based on the forward energy ($E_F$) deposited in the PSD, which
allows to obtain the fraction of total inelastic cross-section by comparing the $E_F$ deposition with 
the predictions of the Wounded Nucleon Model.

Data taking with inserted and removed target was
alternated in order to calculate a data-based correction for interactions with
the material surrounding the target. 
Further details on the experimental
setup, beam and the data acquisition can be found in~Ref.~\cite{Facility}.

\section{Reactions and the methods of particle identification}
This paper presents results from inelastic p+p and centrality selected $^7$Be+$^9$Be interactions and compares them with 
data published by other experiments, including results of Pb+Pb, and Au+Au collisions.

The following methods of particle identification are used for the analyses described in the following sections.
\begin{itemize}
\item The $h^-$ method \cite{PionPaper} is used for calculation of $\pi^-$ spectra. For this method,
all negatively charged particles are treated as $\pi^-$ and a simulation-determined correction factor
is applied to account for the small contamination by K$^-$, $\bar{p}$ and products of weak decays (feed-down).
\item Specific energy loss (dE/dx) within the active volume of the TPCs is used for identification
of charged particles i.e. $p$, $\bar{p}$, K$^\pm$, and $\pi^\pm$ employing a statistical method. 
The dE/dx measurements are binned in momentum $p$ and transverse momentum $p_T$ of the particles. 
For each bin a fit is performed to a sum of Gauss functions (one for each particle type). 
Using the fitted functions, each track is assigned 
a probability of being a particle of given type. The summed probabilities provide the
respective particle multiplicities. 
\item Time of flight (tof) measurement was combined with dE/dx information, allowing to separate 
different kinds of particles in the mid-rapidity region, where the cross-over in specific energy 
loss is the most significant. The square of the particle mass $m^2$ is calculated 
from the particle track length, time of flight, and momentum. The 2-dimensional particle distribution 
in dE/dx and $m^2$ in ($p$, $p_T$) bins is then fitted to 2-dimensional Gaussian distributions and the multiplicities of different
particle types are obtained as above.
\item $\Lambda$ hyperons are identified using a special pattern finding algorithm 
to find the V$^0$-topology characteristic for their decay. 
The invariant mass of the outgoing particle pair is calculated under the assumption of $p$ and $\pi^-$ masses.
The resulting invariant mass distribution for the appropriate phase-space bin is fitted by a sum of a Lorentz function (signal)
and Chebyshev polynomial of 2$^{nd}$ order (combinatorial background) to obtain the $\Lambda$ yield.
Acceptance, feed-down, detector and reconstruction efficiency are corrected using simulations. 
\end{itemize}
%
%\begin{figure}
%\begin{center}
%\resizebox{0.62\textwidth}{!}{
%  \includegraphics{Pictures/dedx.png}
%}
%\resizebox{0.37\textwidth}{!}{
%  \includegraphics{Pictures/tofVsDedx.png}
%}
% %\resizebox{0.42\textwidth}{!}{
% %  \includegraphics{Pictures/lambdaFit.pdf}
% %}
%\end{center}
%\caption{D.}
%\label{fig:idMethods} % Give a unique lebel
%\end{figure}

\section{Results for p+p interactions}
\subsection{Observables used as signatures of the onset of deconfinement}

The main part of the NA61/SHINE program focuses on the observables 
%which are predicted by the Statistical Model of the Early Stage (SMES)~\cite{SMES} 
sensitive to the phase transition between hadron gas and the quark-gluon plasma (QGP)
and the presence of a critical point (CP) of strongly interacting matter. 
High quality data from p+p reactions presented in this section are a 
mandatory reference for the measurements involving heavy ions.

\begin{figure}
\begin{center}
\resizebox{0.45\textwidth}{!}{
  \includegraphics{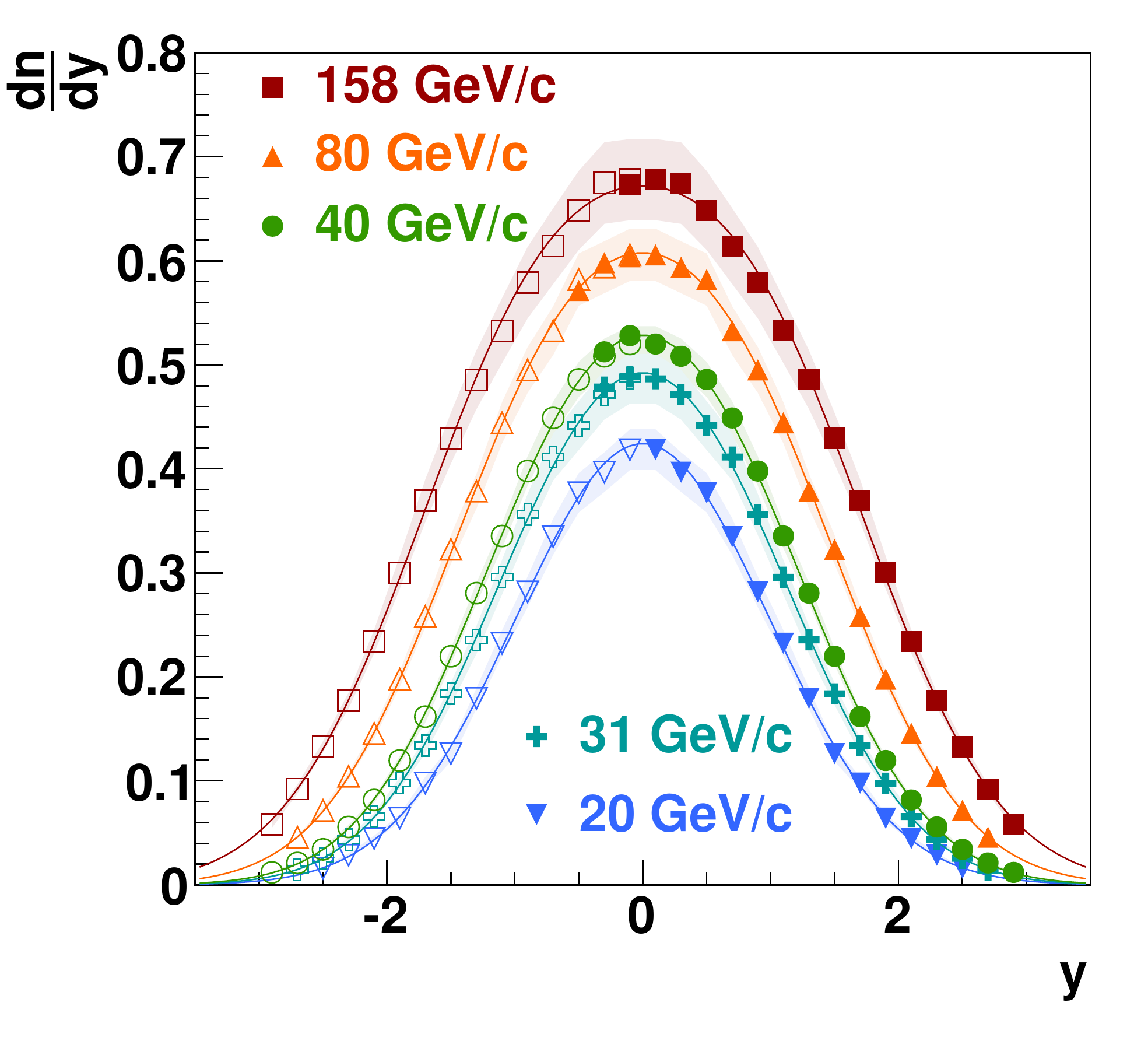}
}
\resizebox{0.45\textwidth}{!}{
  \includegraphics{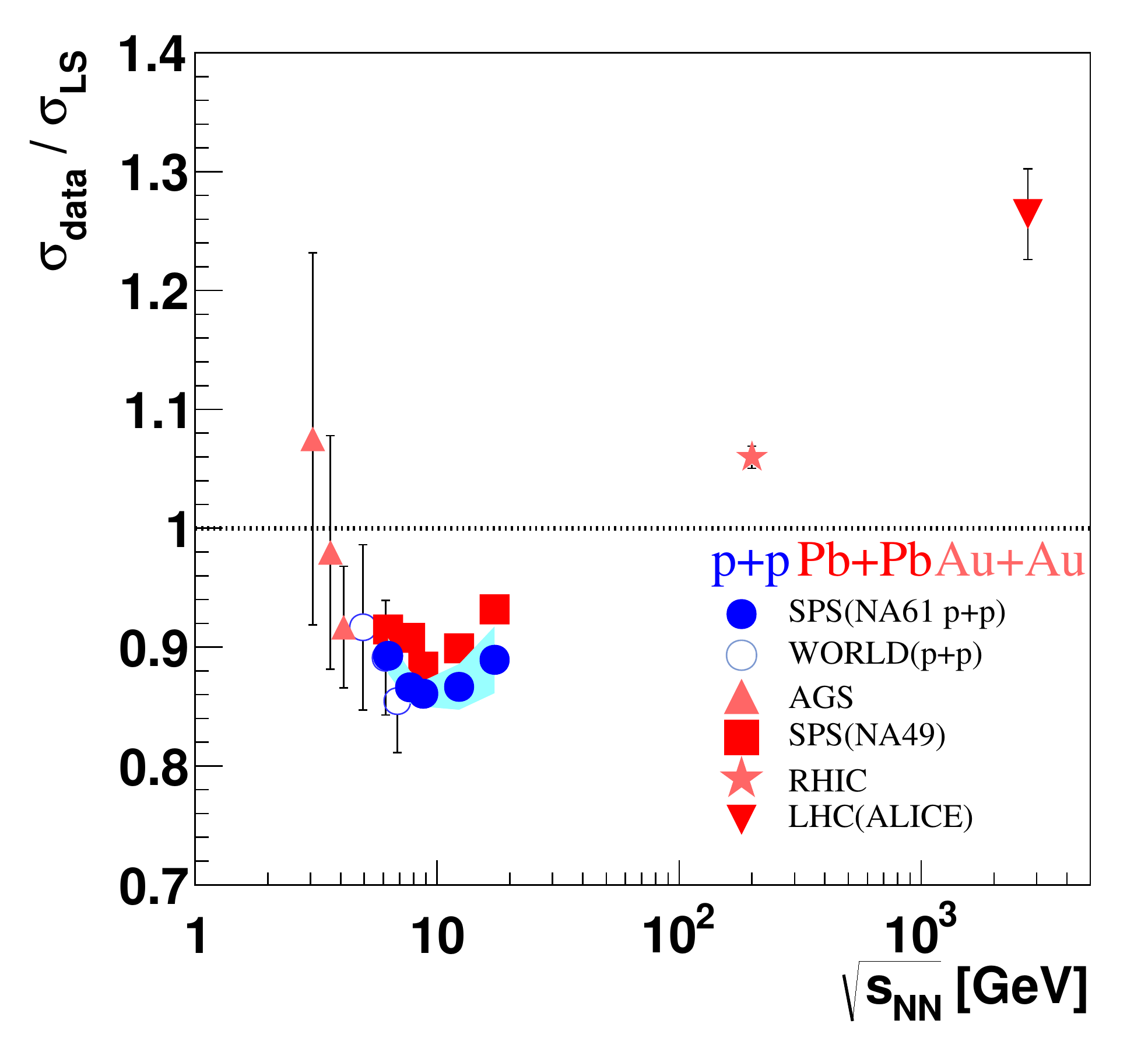}
}
\end{center}
\caption{\textit{Left:} rapidity dependence of $\pi^-$ yields in inelastic p+p interactions 
for momenta ranging from 20 to 158~\GeVc fitted to a sum of two identical Gauss functions
symmetrically displaced from mid-rapidity. \textit{Right:} dale in the width of the rapidity distribution 
normalised to $\sigma_{LS}$ (from the Landau-Shuryak hydrodynamic model~\cite{Landau,Shuryak} 
observed at $\sqrt{s_{NN}}\approx10$~\GeV. The corresponding minimum of 
the sound velocity is generally interpreted as softening of the EoS due to 
creation of a mixed phase at the transition energy from hadrons to partonic matter \cite{SMES}. 
Interestingly, this signature is found not only for heavy ion collisions \cite{Ten,Eleven,Twelve}, 
but also for p+p reactions. The results are not corrected for isospin effects.}
\label{fig:dale} % Give a unique lebel
\end{figure}

One of the signatures predicted for the phase transition is a clear minimum in
the energy dependence of the velocity of sound $c_s$. This effect is due 
to the softening of the equation of state (EoS) in the mixed state system 
of a 1st order phase transition.
As the width $\sigma$ of the rapidity distribution can be expressed by $c_s$~\cite{Landau}
\begin{equation}
\sigma^2=\frac{8}{3}\frac{c_s^2}{1-c_s^4}\ln{\left(\sqrt{s_{NN}}/2m_N\right)}~,
\end{equation}
the minimum should be directly visible for the width of $\pi^-$ rapidity spectra (Fig.~\ref{fig:dale}~{\it left}). 
The minimum is expected only in the case of a transition to the QGP in heavy-ion collisions.
Interestingly, such a structure emerges also in p+p collisions (Fig.~\ref{fig:dale}~{\it right}).

%\begin{figure}
%\begin{center}
%\resizebox{0.45\textwidth}{!}{
%  \includegraphics{Pictures/szymon/AAblast_pp.pdf}
%}
%\resizebox{0.45\textwidth}{!}{
%  \includegraphics{Pictures/szymon/AAblast_PbPb.pdf}
%}
%\end{center}
%\caption{D.}
%\label{fig:blasta} % Give a unique lebel
%\end{figure}

\begin{figure}
\begin{center}
\resizebox{0.45\textwidth}{!}{
  \includegraphics{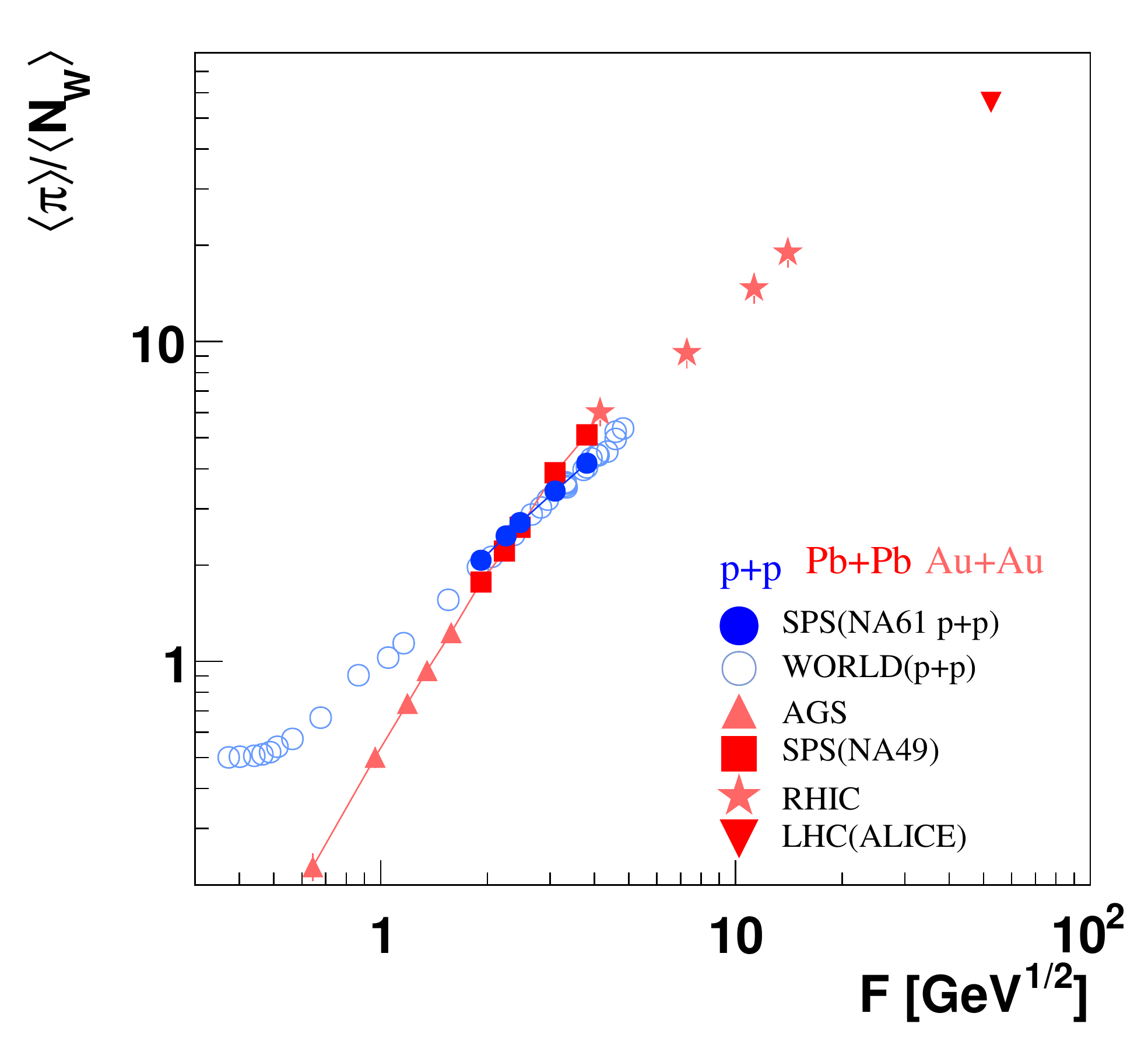}
}
\resizebox{0.45\textwidth}{!}{
  \includegraphics{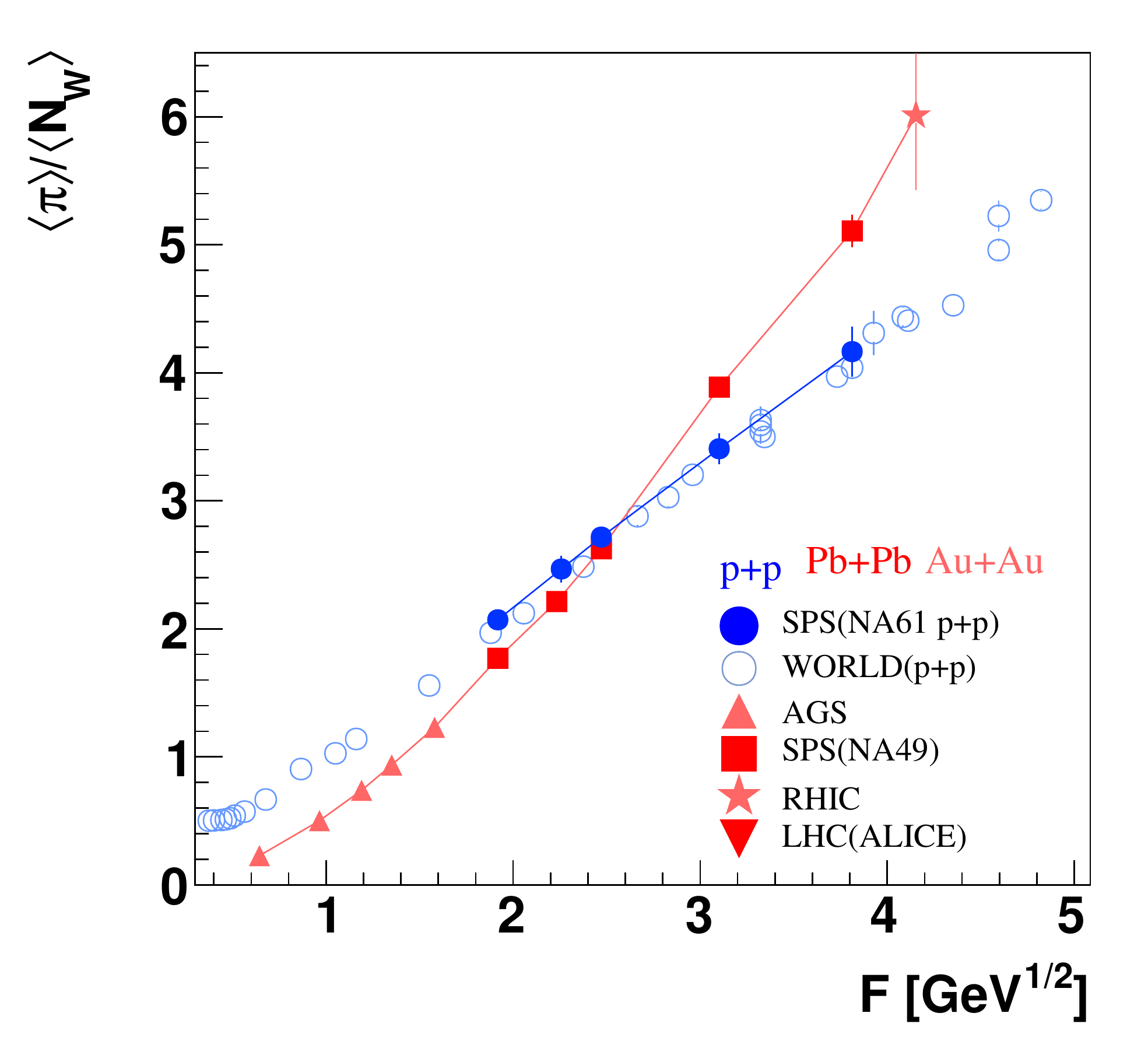}
}
\end{center}
\caption{Dependence of $\pi$ multiplicity on the energy variable $F$~(see text). Also plotted is a
compilation of world data based on Ref.~\cite{Eleven,Thirteen} including the preliminary \NASixtyOne results
on p+p interactions (solid circles). Double-logarithmic scale (\textit{left}), linear scale (\textit{right}).}
\label{fig:kink} % Give a unique lebel
\end{figure}

A signature predicted by the Statistical Model of the Early Stage (SMES)~\cite{SMES} 
is the ratio of entropy (measured by the multiplicity of pions) to the number of 
wounded nucleons (interacting projectile and target nucleons). 
The SMES model predicts a linear increase with the energy variable $F=\frac{(\sqrt{s}-2m_p)^{3/4}}{\sqrt{s}^{1/4}}$~\cite{Fermi} 
for the situation without the phase transition, while the creation of the QGP
results in an increase of the slope ('the kink'), because the produced entropy increases due to 
the activation of the partonic degrees of freedom. 
The~measurement of pion multiplicities in inelastic p+p interactions at SPS energies and the world 
data (see Fig.~\ref{fig:kink}) suggest that in fact for $F>1$~\GeV$^{1/2}$
pion production rises linearly, while the data for Pb+Pb, and Au+Au show a significant steepening
of the rate of increase of pion production between 40 and 80\AGeVc.

\begin{figure}
\begin{center}
\resizebox{0.45\textwidth}{!}{
  \includegraphics{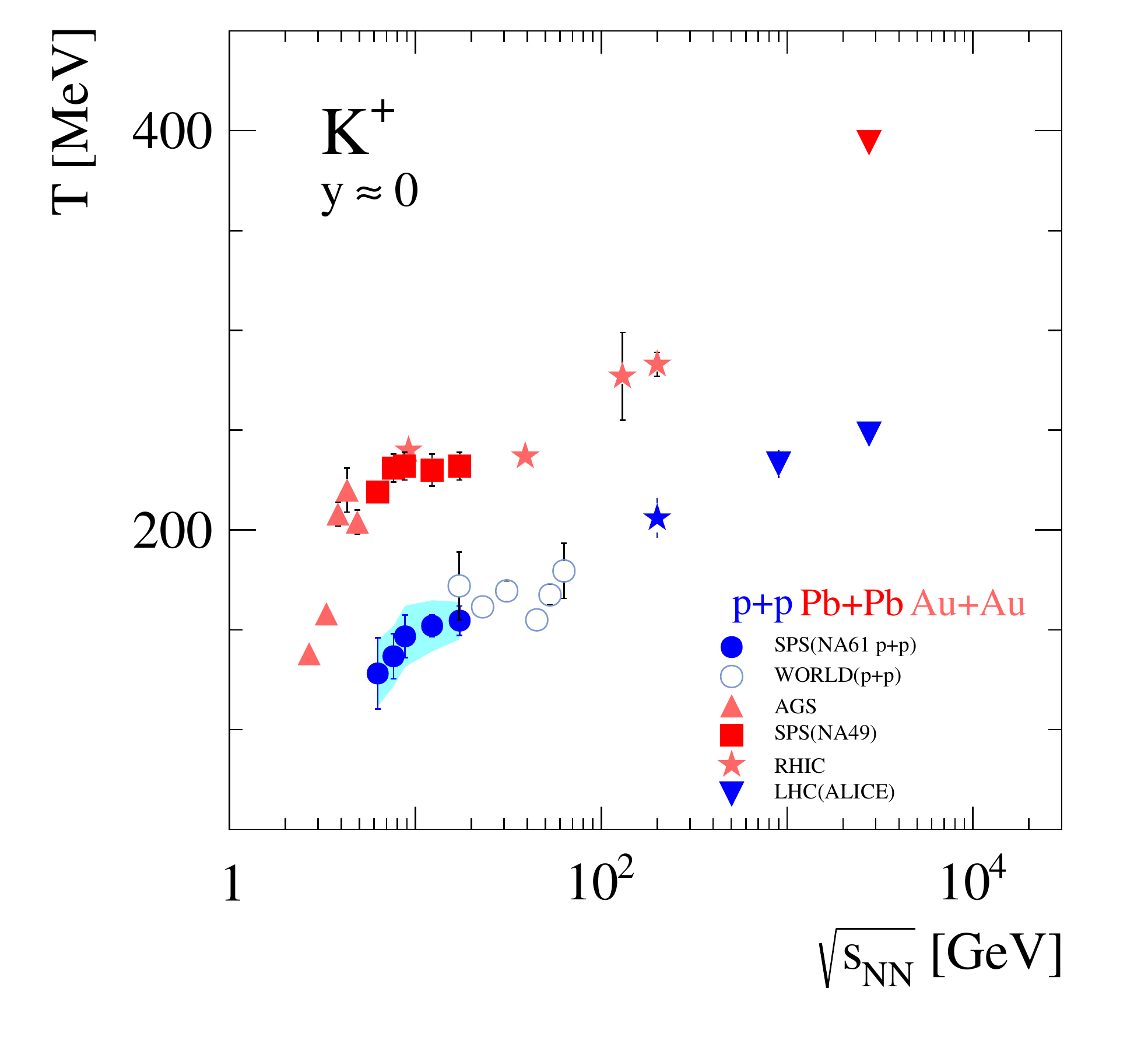}
}
\resizebox{0.45\textwidth}{!}{
  \includegraphics{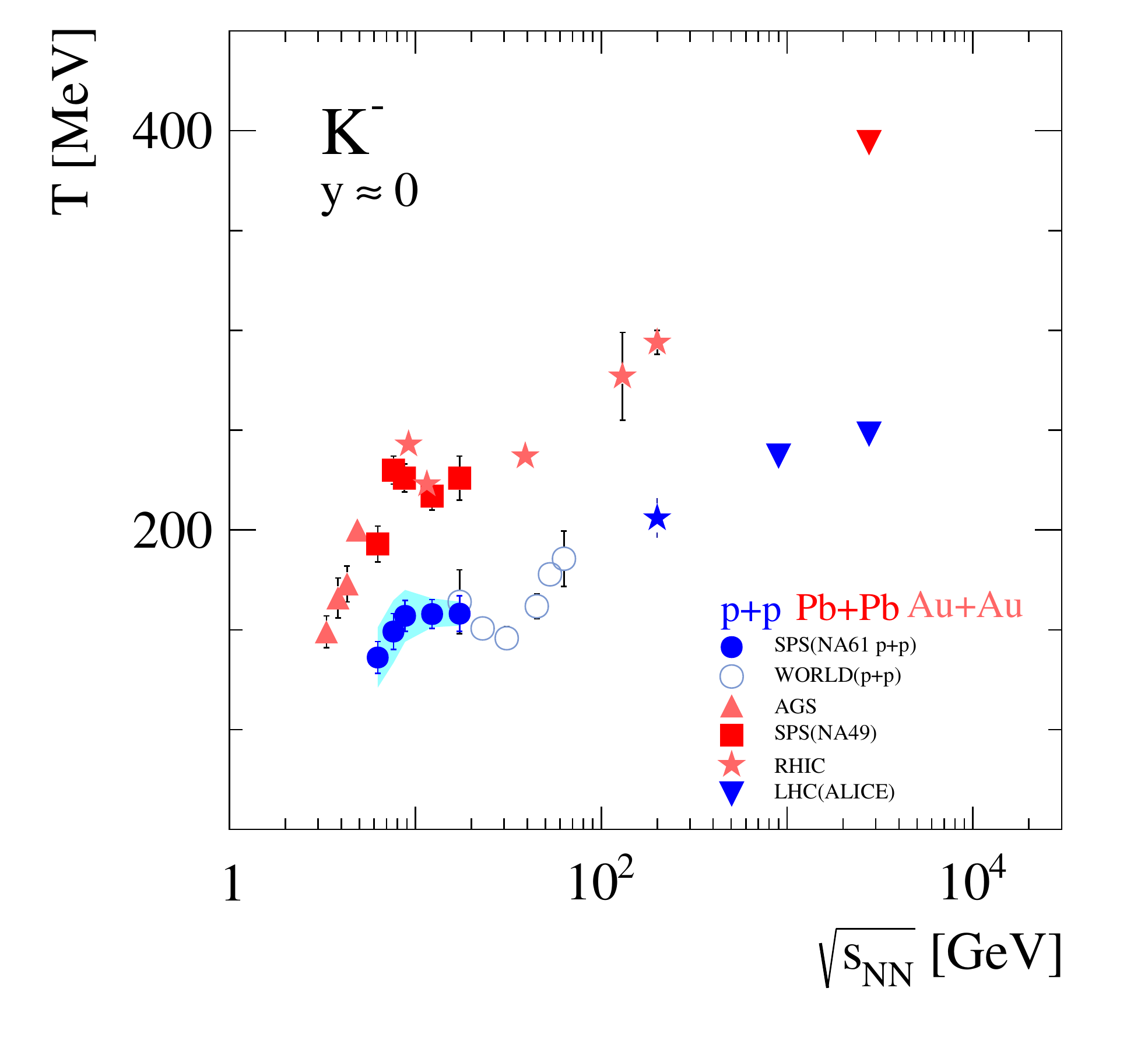}
}
\end{center}
\caption{Energy dependence of inverse slope parameter $T$ of kaon transverse mass spectra 
($K^+$~(\textit{left}), $K^-$~(\textit{right})) showing preliminary \NASixtyOne results 
on p+p interactions (solid circles) and a compilation of world data from Ref.~\cite{Fourteen,Seventeen,Eighteen,Nineteen}.}
\label{fig:step} % Give a unique lebel
\end{figure}

The second SMES-suggested signature of the onset of QGP production analysed by \NASixtyOne was the 
centre-of-mass energy $\sqrt{s_{NN}}$ dependence of the inverse slope parameter $T$
of the transverse mass $m_T$ distributions for $K^{\pm}$ at mid-rapidity. 
A stationary behaviour (the 'step') was predicted in the mixed-phase region and actually observed in central Pb+Pb collisions. 
Surprisingly, the \NASixtyOne results from inelastic p+p interactions (Fig.~\ref{fig:step}) also exhibit a 'step'
structure like that found in central Pb+Pb collisions.

\begin{figure}
\begin{center}
\resizebox{0.45\textwidth}{!}{
  \includegraphics{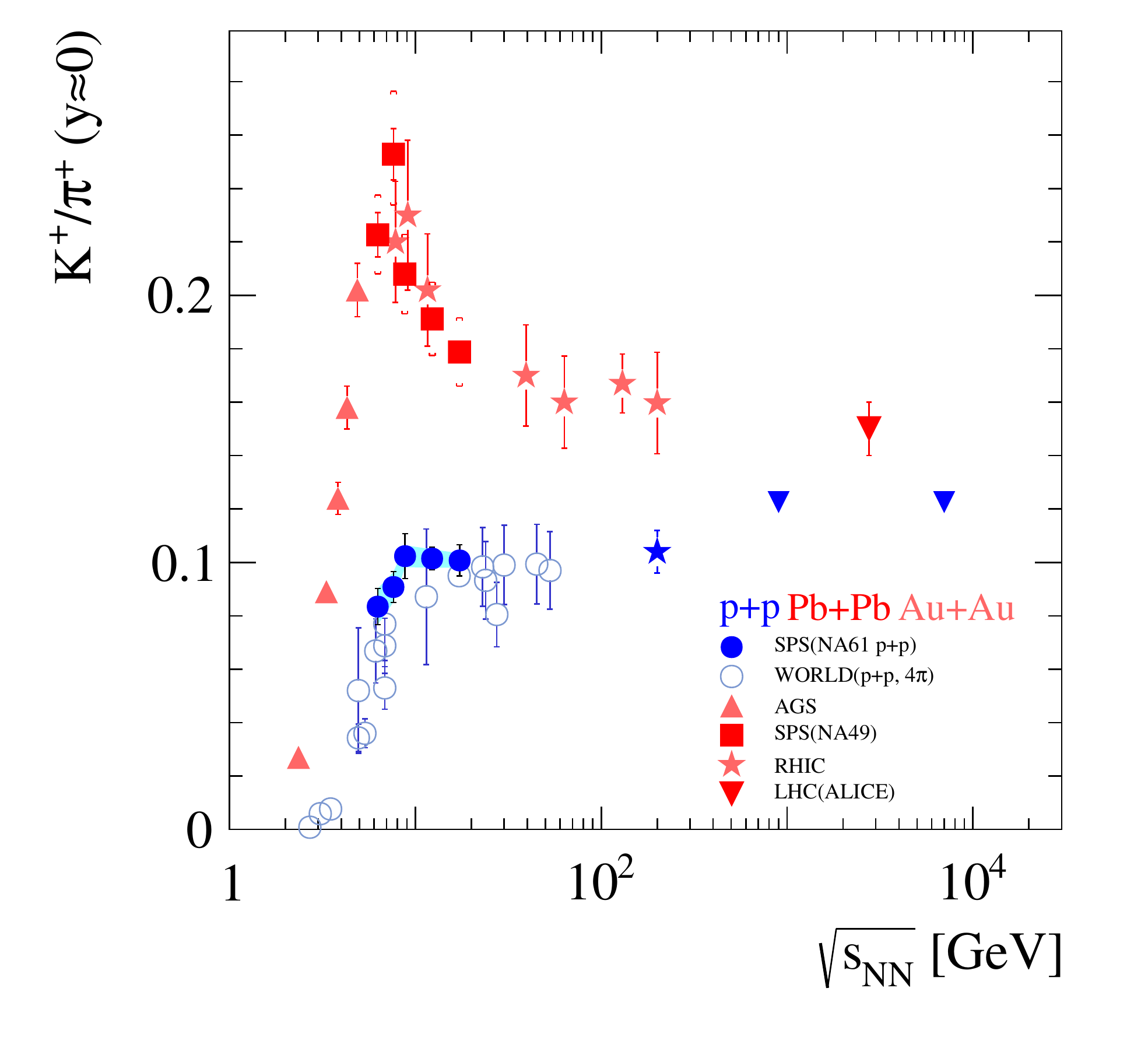}
}
\resizebox{0.45\textwidth}{!}{
  \includegraphics{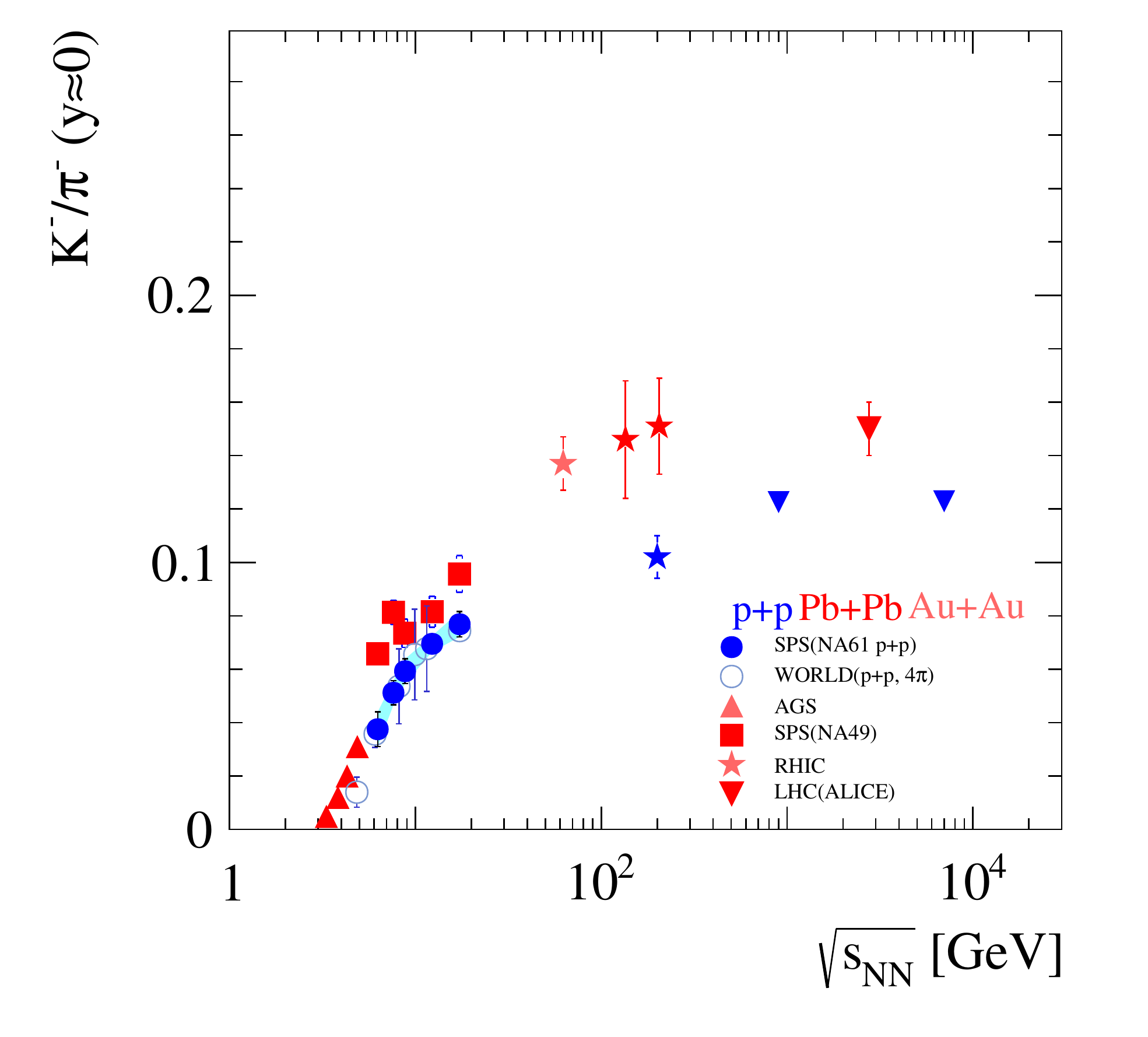}
}
\end{center}
\caption{\textit{Left}: the energy dependence of the $K^+/\pi^+$  ratio in p+p interactions 
changes rapidly at the energy where the 'horn' structure is visible in Pb+Pb. Such a behaviour was 
not predicted in p+p interactions by models, including the Statistical Model of the Early Stage, 
which predicted the 'horn' structure as a signature of the Quark-Gluon Plasma (QGP). 
%The structure in p+p can be reproduced by applying exact strangeness conservation~\cite{StrangenessConservation}.
\textit{Right:} the energy dependence of the $K^-/\pi^-$  ratio in p+p and heavy-ion interactions. 
%The production of $K^-$ depends not only on strangeness production, but also on baryon density.
The energy dependence of the $K^-/\pi^-$ ratio does not reveal the horn structure since
it is not representative of total strangeness production. 
The world data plotted in both panels is taken from the Ref.~\cite{Thirteen,Nineteen,Sixteen,Fifteen,Twenty}.}
\label{fig:horn} % Give a unique lebel
\end{figure}

The most interesting signature is predicted by the SMES in the ratio of strangeness to entropy production. 
The energy dependence of the related ratio $\langle K^+\rangle / \langle \pi^+\rangle$ is expected to show
a rapid increase in the hadron gas phase followed by an abrupt drop at the onset of deconfinement
due to a jump in entropy production and then a smooth decrease due to further increase of entropy. 
This results in the 'horn' structure in heavy-ion collisions, which is not expected
in the reference p+p data, as the transition to the QGP is improbable there. Actually, 
a step-like structure (precursor of the 'horn') is visible in inelastic 
p+p interactions (Fig.~\ref{fig:horn}) motivating a more thorough theoretical study with
incorporation of strict strangeness conservation \cite{StrangenessConservation}.

\subsection{Strange neutral particles - $\Lambda$}
\begin{figure}
\begin{center}
\resizebox{0.45\textwidth}{!}{
  \includegraphics{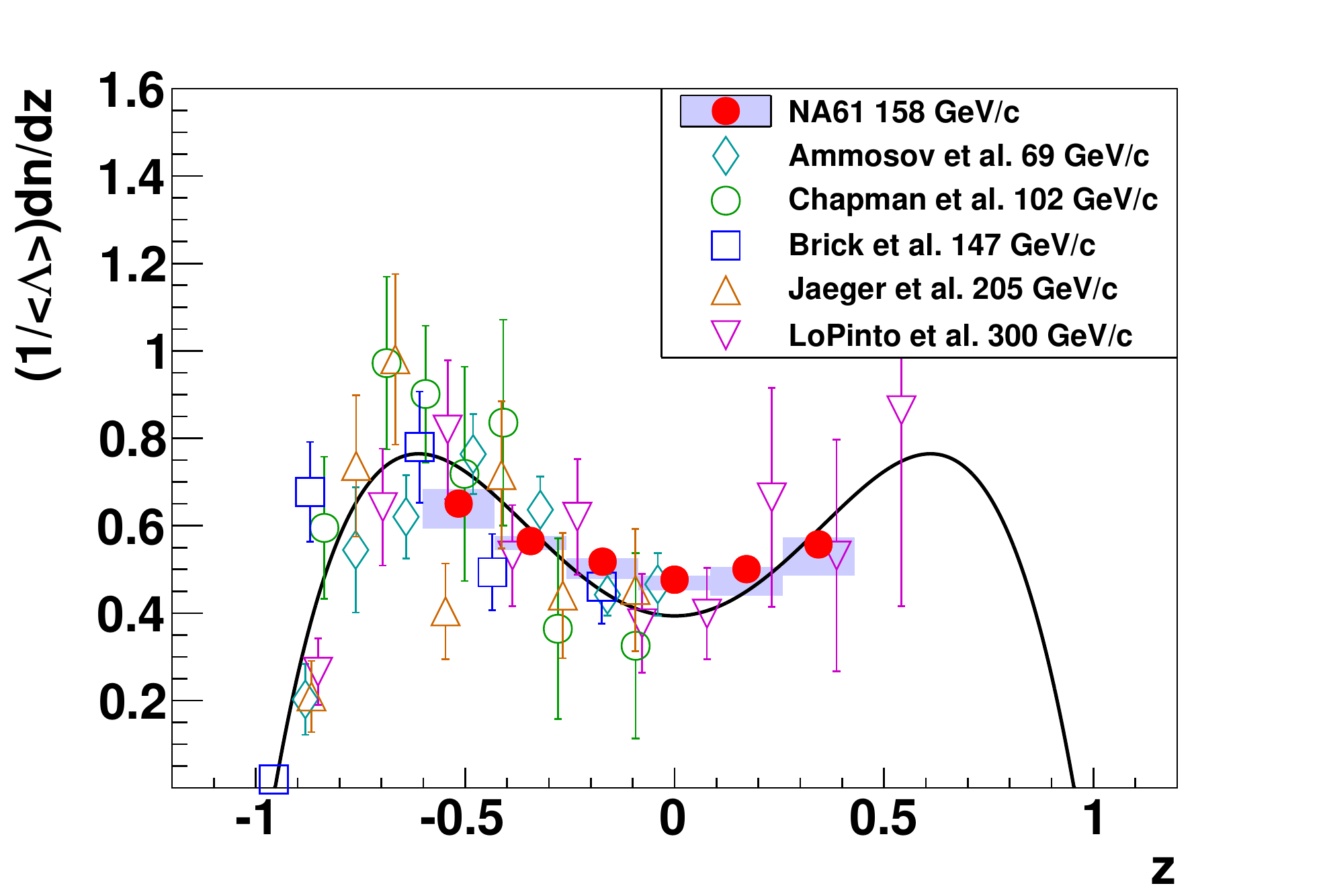}
}
\resizebox{0.45\textwidth}{!}{
  \includegraphics{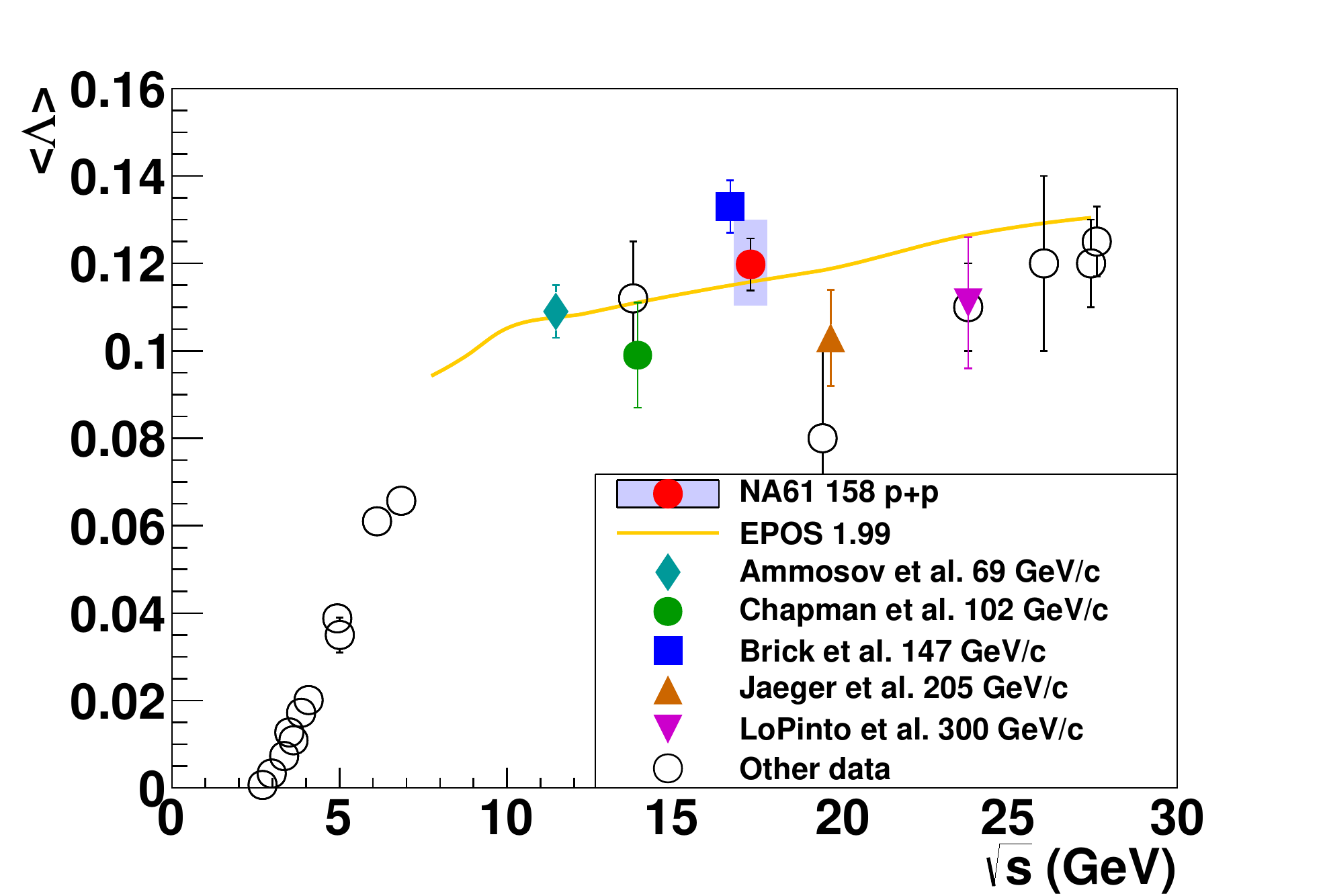}
}
\end{center}
\caption{\textit{Left:} The $\Lambda$ yield as function of scaled rapidity $z=y/y_{beam}$
and normalised to unity in inelastic p+p interactions measured by 
\NASixtyOne  and selected bubble-chamber
experiments~\cite{Ammosov,Chapman,Brick,Jaeger,LoPinto}.
The symmetric polynomial of 4$^{th}$ order used for estimation of the
systematic uncertainty of $\Lambda$ total multiplicity is plotted to guide the eye.
\textit{Right:} Collision energy dependence of total multiplicity of $\Lambda$ 
hyperons produced in inelastic p+p interactions.
Full symbols indicate
bubble chamber results,
the solid red dot shows the \NASixtyOne result.
Open symbols depict the remaining world data~\cite{Sixteen}.
The prediction of the \EposLong~\cite{EPOS} model is shown by the curve. The
systematic uncertainty of the \NASixtyOne result is indicated
by the shaded bar.}
\label{fig:lambda} % Give a unique lebel

\end{figure}

The rapidity spectrum of $\Lambda$ hyperons from \NASixtyOne is compared in Fig.~\ref{fig:lambda} (\textit{left}) to
results from five bubble-chamber experiments which measured p+p interactions
at beam momenta close to 158~\GeVc.
The experiments published data for the backward hemisphere,
however, with rather small statistics~\cite{Ammosov,Chapman,Brick,Jaeger,LoPinto}
and correspondingly large uncertainties. 
To account for the difference in beam momentum
the spectra are shown in terms of the scaled rapidity $z=y/y_{beam}$
and were normalised to unity in order to compare the shapes.

Though the statistical error and the systematic uncertainty of the \NASixtyOne
measurement is much smaller than
for the other experiments, and the results are consistent with all the
datasets used for the comparison, the general tendency obtained
by fitting a symmetric polynomial of 4$^{th}$ order does not describe well
the \NASixtyOne data. On the other hand, the result of Brick {\it et al.}
for which the beam momentum (147~\GeVc) differs the least from the
\NASixtyOne momentum, shows the best agreement.

The estimated total multiplicity of $\Lambda$ hyperons produced in inelastic p+p interactions at 158~\GeVc
is compared in Fig.~\ref{fig:lambda} (\textit{right}) with the world data~\cite{Sixteen} 
as well as with predictions of the \EposLong model in its validity range.
A steep rise in the threshold region is followed by a more gentle increase
at higher energies that is well reproduced by the \EposLong model.

\subsection{Two-particle $\Delta\eta\Delta\phi$ correlations}
The correlation between charged particles in centre-of-mass pseudo-rapidity $\eta$ and azimuthal angle $\phi$
was measured by the following 2-particle correlation function: 
\begin{equation}
C(\Delta\eta,\Delta\phi)=\frac{N^{pairs}_{mixed}}{N^{pairs}_{single}}\frac{S(\Delta\eta,\Delta\phi)}{M(\Delta\eta,\Delta\phi)}~,
\end{equation}
where $\Delta\eta$ and $\Delta\phi$ are the rapidity and azimuthal angle difference of the particles and
\begin{equation}
S(\Delta\eta,\Delta\phi)=\frac{d^2N^{pairs}_{single}}{d\Delta\eta,\Delta\phi}~,
\end{equation}
and
\begin{equation}
M(\Delta\eta,\Delta\phi)=\frac{d^2N^{pairs}_{mixed}}{d\Delta\eta,\Delta\phi}~.
\end{equation}
Here $N^{pairs}_{single}$ denotes the number of pairs in the events and $N^{pairs}_{mixed}$ the number of pairs of tracks
taken from different events, the latter representing the uncorrelated reference. 

The $\Delta\phi$ range is folded, i.e. for $\Delta\phi$ larger than $\pi$ its value is recalculated as $2\pi-\Delta\phi$. 
Detector effects were corrected using simulations and were found to be small. 

\begin{figure}
\begin{center}
\resizebox{0.76\textwidth}{!}{
  \includegraphics{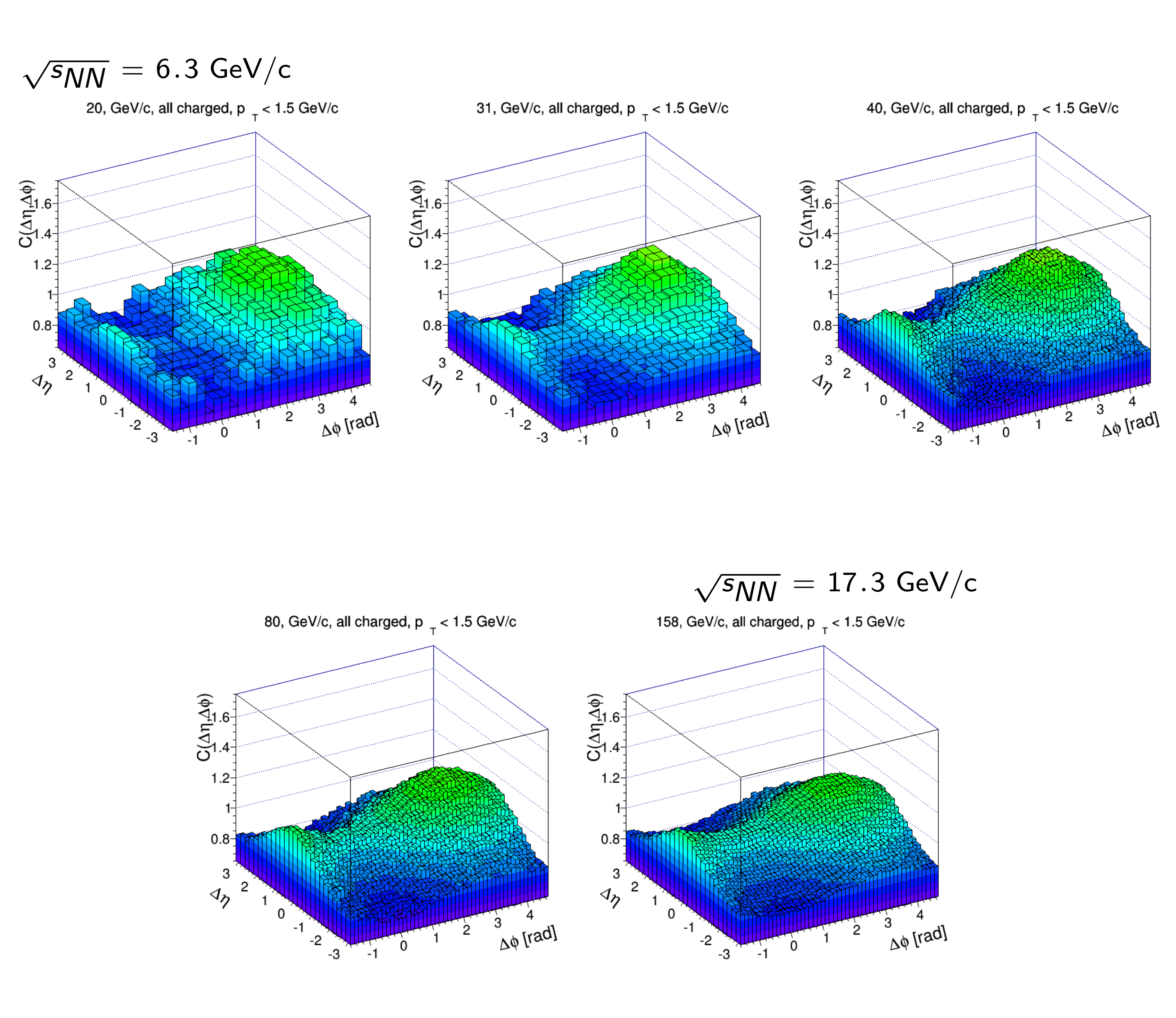}
}
\resizebox{0.23\textwidth}{!}{
  \includegraphics{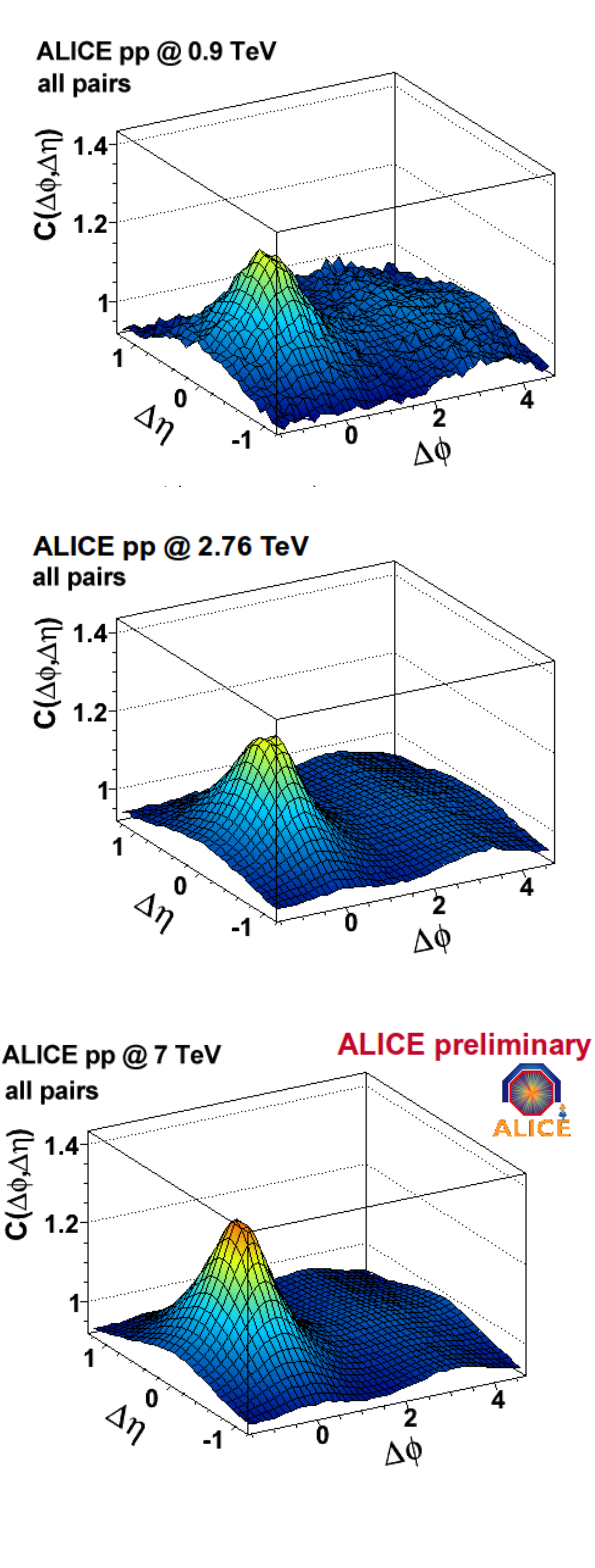}
}
\end{center}
\caption{\textit{Left:} Correlation functions $C(\Delta\eta,\Delta\phi)$ measured by NA61/SHINE show a maximum at ($\Delta\eta,\Delta\phi$)=(0,$\pi$) probably due to resonance decays and momentum conservation.
In comparison with the measurement performed at ultra-relativistic energies by ALICE~\cite{Janik} (\textit{right}), 
the NA61/SHINE results show a stronger enhancement for $\Delta\phi\approx\pi$ but no 'jet peak'.} %\\at $\Delta\phi\approx 0$ }
\label{fig:coincidences} % Give a unique lebel
\end{figure}

The results in Fig. \ref{fig:coincidences} (\textit{left}) show a clear enhancement at ($\Delta\eta,\Delta\phi$)=(0,$\pi$) 
most likely due to resonance decays and momentum conservation, 
and a weaker enhancement at ($\Delta\eta,\Delta\phi$)=(0,0) probably due to Coulomb interactions and quantum statistics. 

\section{Results of $^7$Be+$^9$Be interactions}
\subsection{Inelastic $^7$Be+$^9$Be cross section}
\begin{figure}
\begin{center}
\resizebox{0.7\textwidth}{!}{
  \includegraphics{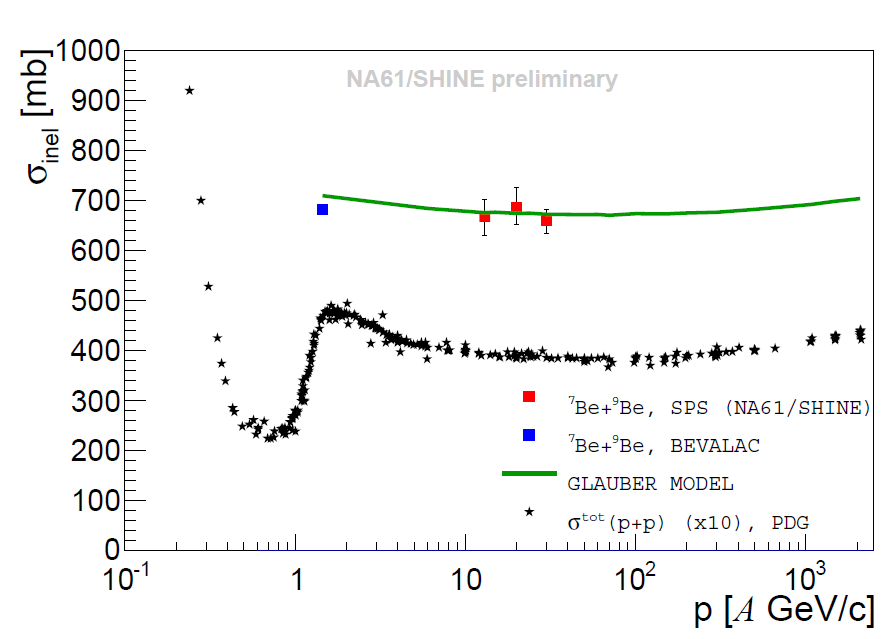}
}
\end{center}
\caption{Beam momentum dependence of the total inelastic cross section for $^7$Be+$^9$Be interactions. The \NASixtyOne
data (red squares), together with the LBNL result (blue square)~\cite{Lbnl} establish the energy dependence of the cross section. 
Calculations using the Glissando model \cite{Glissando} are shown by the green curve for comparison. }
\label{fig:crosssection} % Give a unique lebel
\end{figure}

The inelastic cross section for $^7$Be+$^9$Be interactions was determined for 13$A$, 20$A$, and 30\AGeVc using the pulse-height distributions of the S4 counter - 
a round scintillator counter of 2~cm in diameter (see Fig.~\ref{fig:na61}). 
%It was placed at the beam position between \VTPCOne, and \VTPCTwo (Fig.~\ref{fig:na61}).
The values measured by the \NASixtyOne experiment (Fig.~\ref{fig:crosssection})~\cite{Weimer} are in good agreement with an earlier measurement at lower momentum \cite{Lbnl}, as well as with
the predictions of the Glauber-based Glissando model~\cite{Glissando}. 

\subsection{Rapidity distributions}
\begin{figure}
\begin{center}
\begin{minipage}[t]{0.32\textwidth}\includegraphics[width=0.99\textwidth]{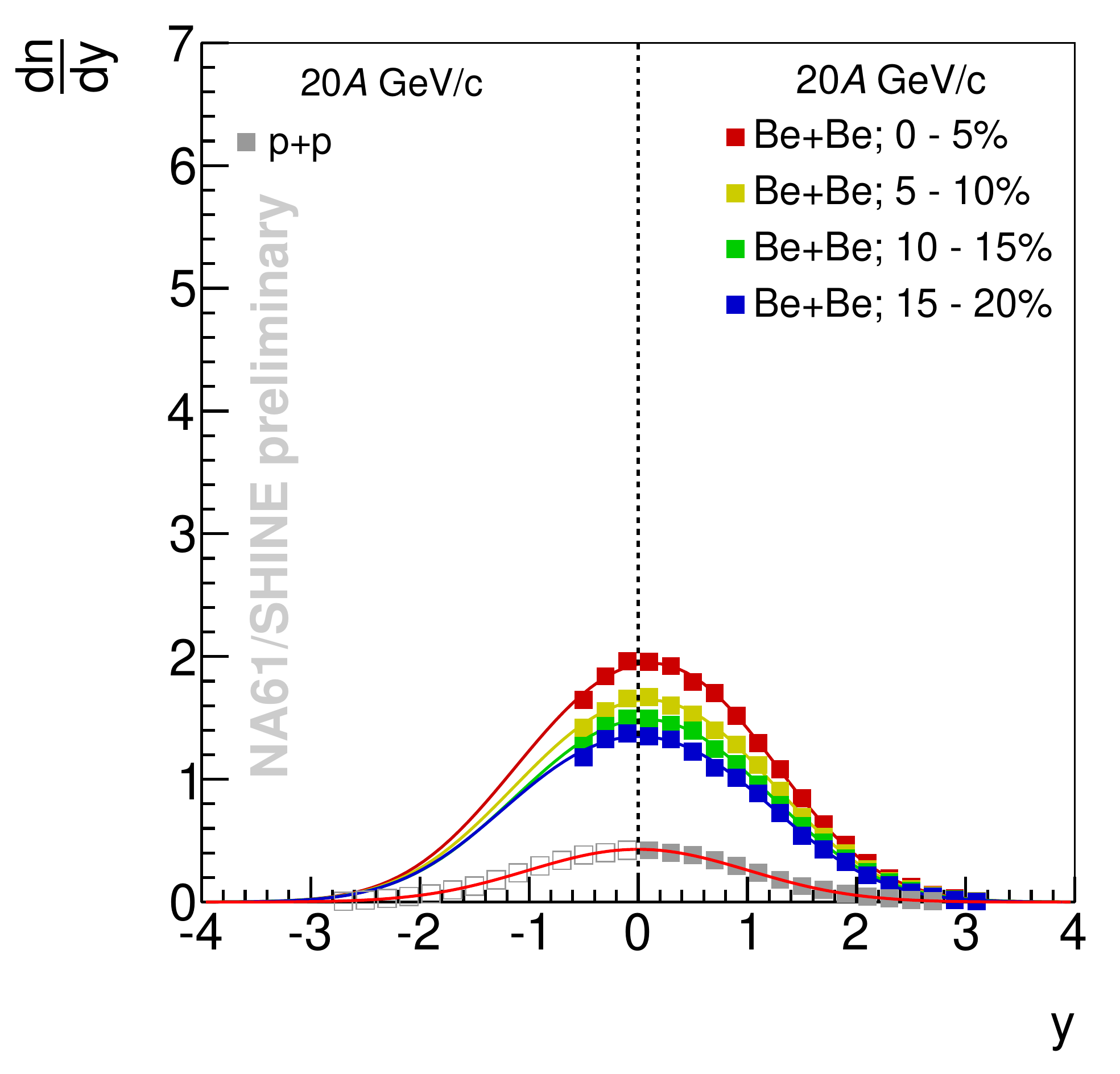}\end{minipage}
\begin{minipage}[t]{0.32\textwidth}\includegraphics[width=0.99\textwidth]{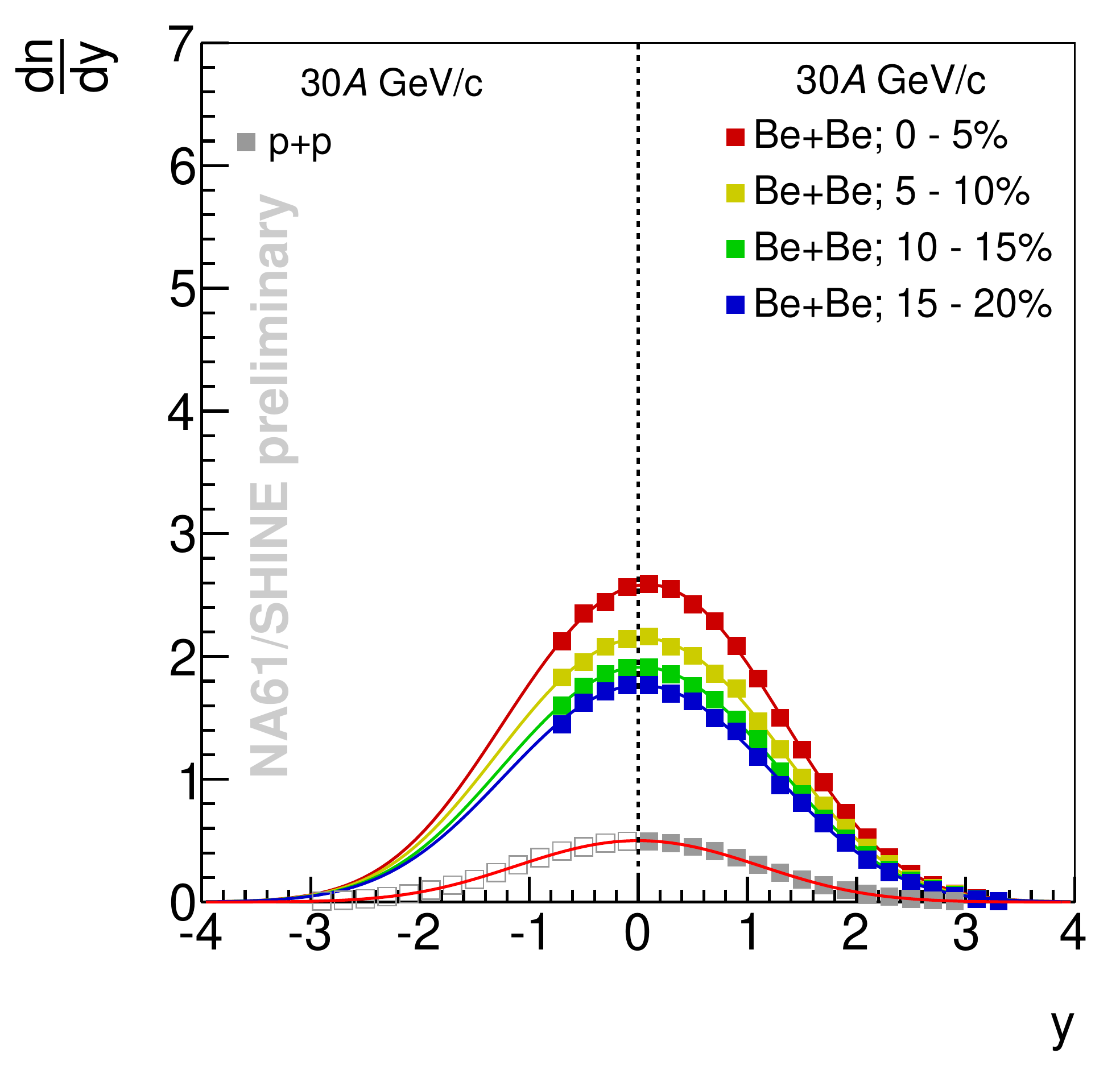}\end{minipage}
\begin{minipage}[b]{0.32\textwidth}\includegraphics[width=0.99\textwidth]{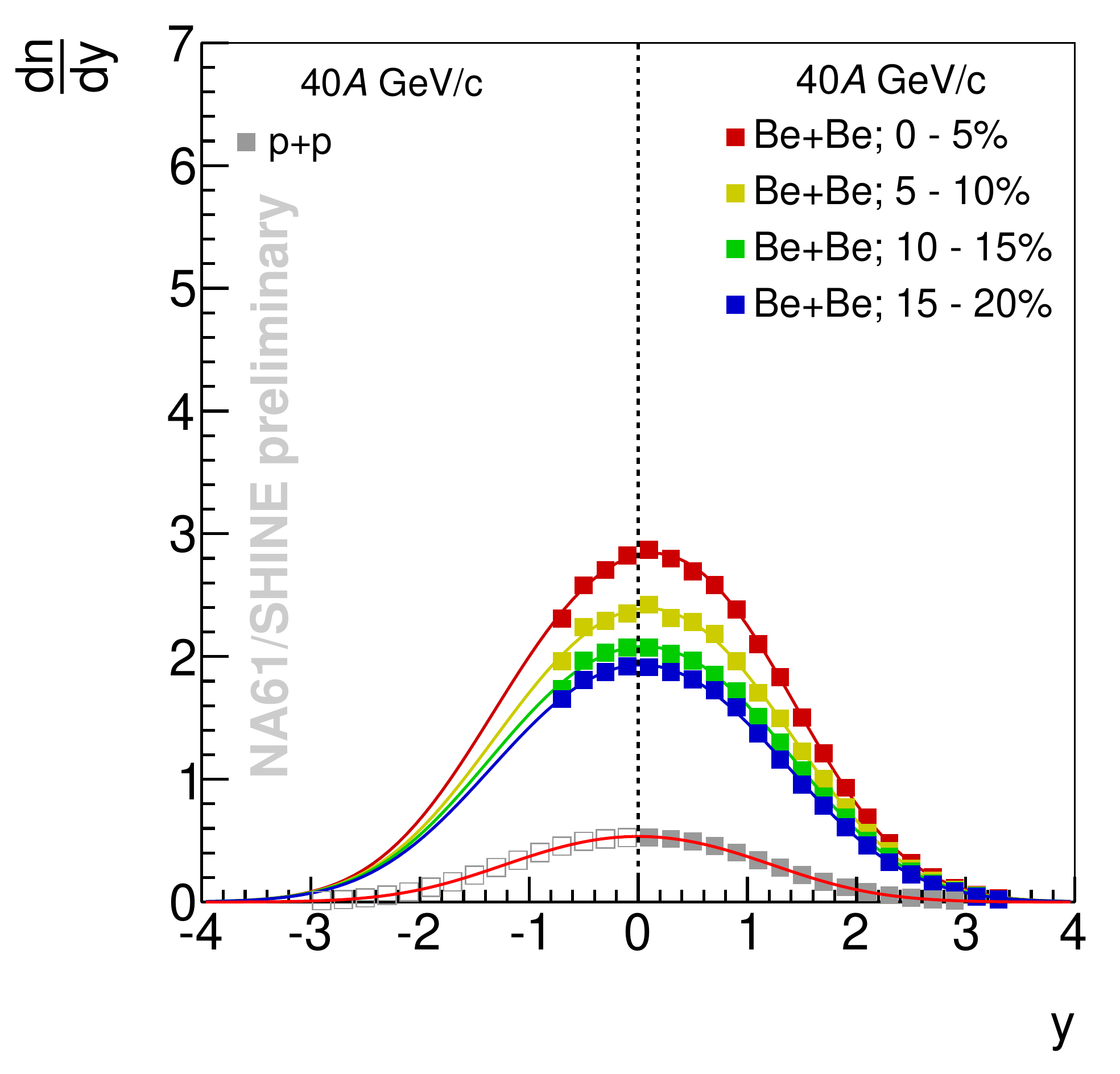}\end{minipage}
\begin{minipage}[b]{0.32\textwidth}\includegraphics[width=0.99\textwidth]{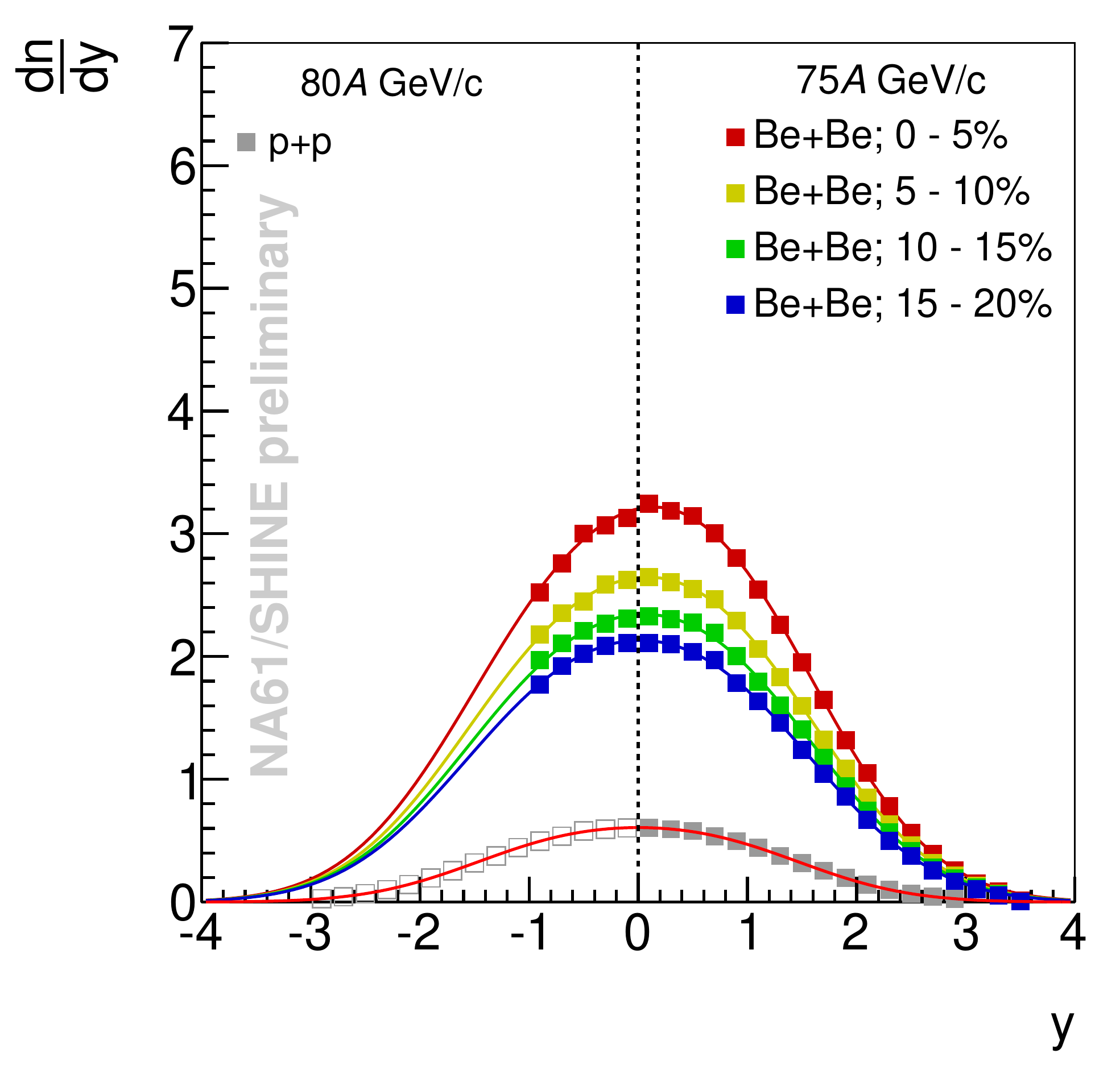}\end{minipage}
\begin{minipage}[b]{0.32\textwidth}\includegraphics[width=0.99\textwidth]{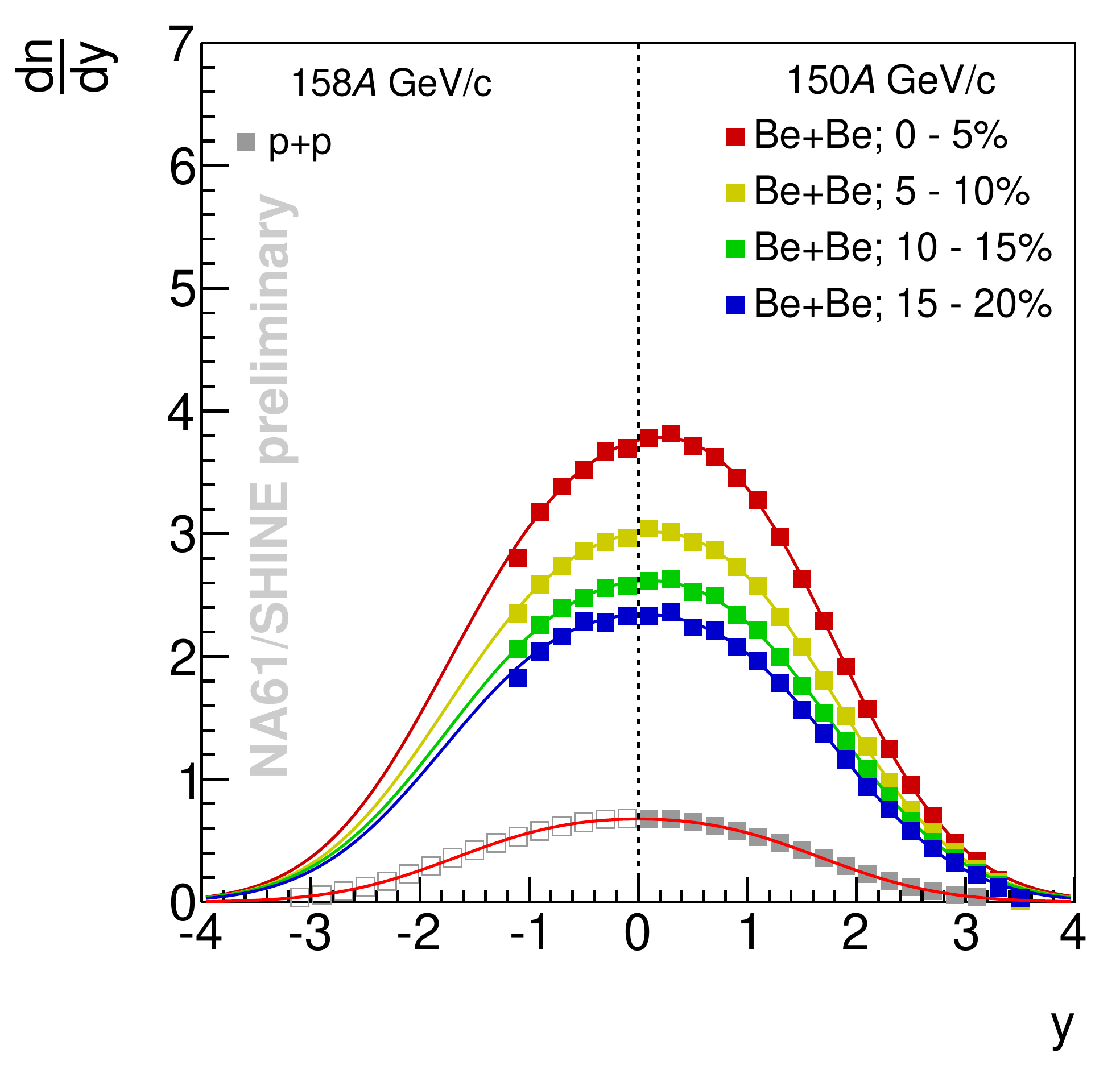}\end{minipage}
\end{center}
\caption{Rapidity spectra of $\pi^-$ in $^7$Be+$^9$Be collisions at 5 beam momenta 20$A$, 30$A$, 40$A$, 75$A$, and 150$A$ \GeVc 
for 4 centrality classes compared
to p+p data \cite{PionPaper} at the nearest measured energy.}
\label{fig:ySpectra} % Give a unique lebel
\end{figure}

The $\pi^-$ rapidity distributions for $^7$Be+$^9$Be collisions at five beam momenta and in four centrality classes 
together with data for inelastic p+p interactions are presented in Fig. \ref{fig:ySpectra}. One observes a small 
asymmetry about mid-rapidity in the rapidity distributions for 
$^7$Be+$^9$Be collisions. There are two effects, which may be responsible for this feature.
On the one hand, there is a mass asymmetry between projectile ($^7$Be) and target ($^9$Be) nuclei, which is expected to enhance particle 
production in the backward hemisphere. On the other hand, the selection of central collisions requires a small number of projectile 
spectators without any restriction imposed on the number of target spectators. For collisions of identical nuclei this would enhance particle 
production in the forward hemisphere. As the two effects compensate each other to a great extent, the asymmetry of the measured spectra 
tends to be relatively small. Figure \ref{fig:ySpectra} shows that the second effect is a little bit stronger.

\subsection{Transverse mass distributions}
\begin{figure}
\begin{center}
\begin{minipage}[t]{0.32\textwidth}\includegraphics[width=0.99\textwidth]{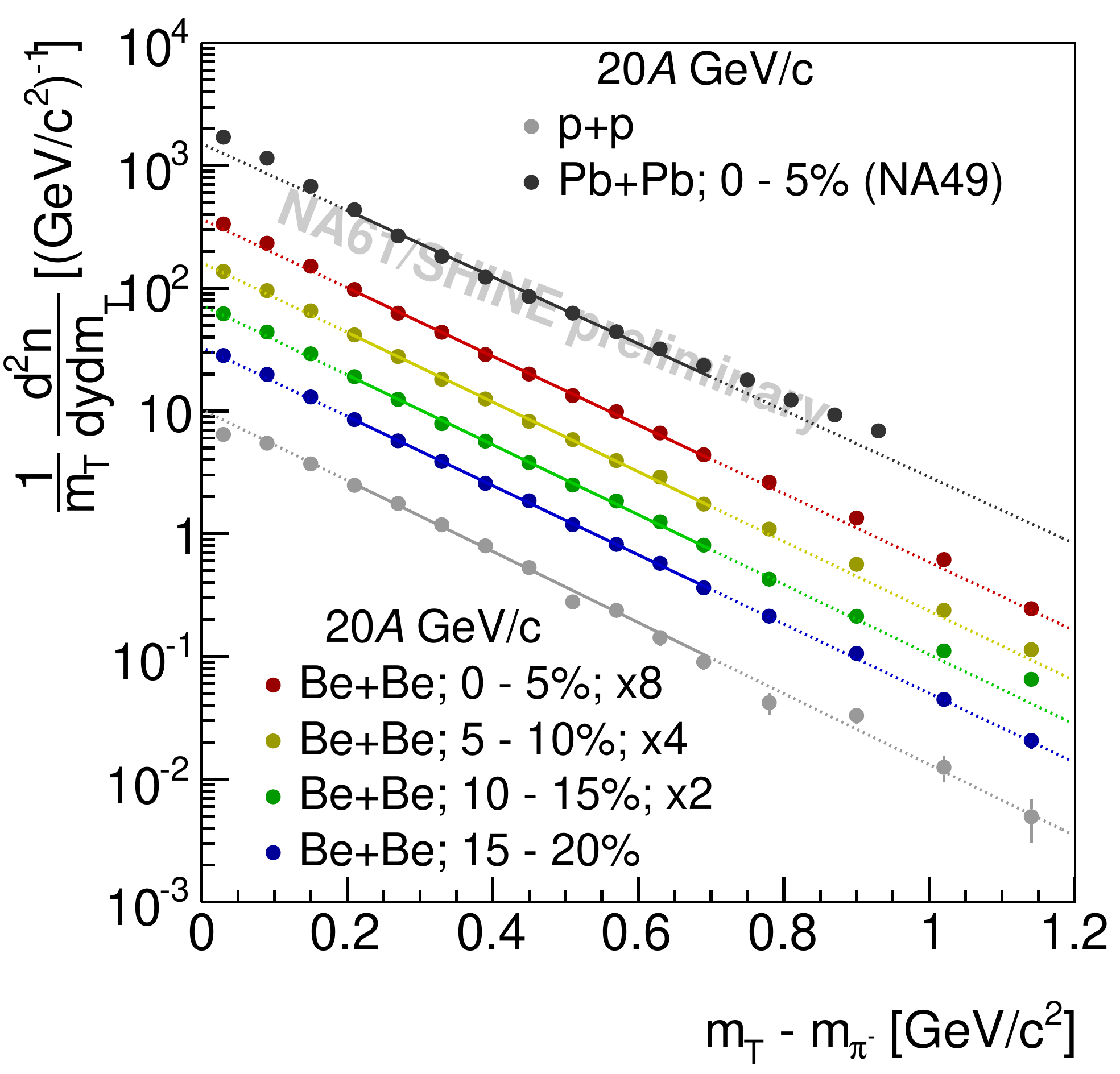}\end{minipage}
\begin{minipage}[t]{0.32\textwidth}\includegraphics[width=0.99\textwidth]{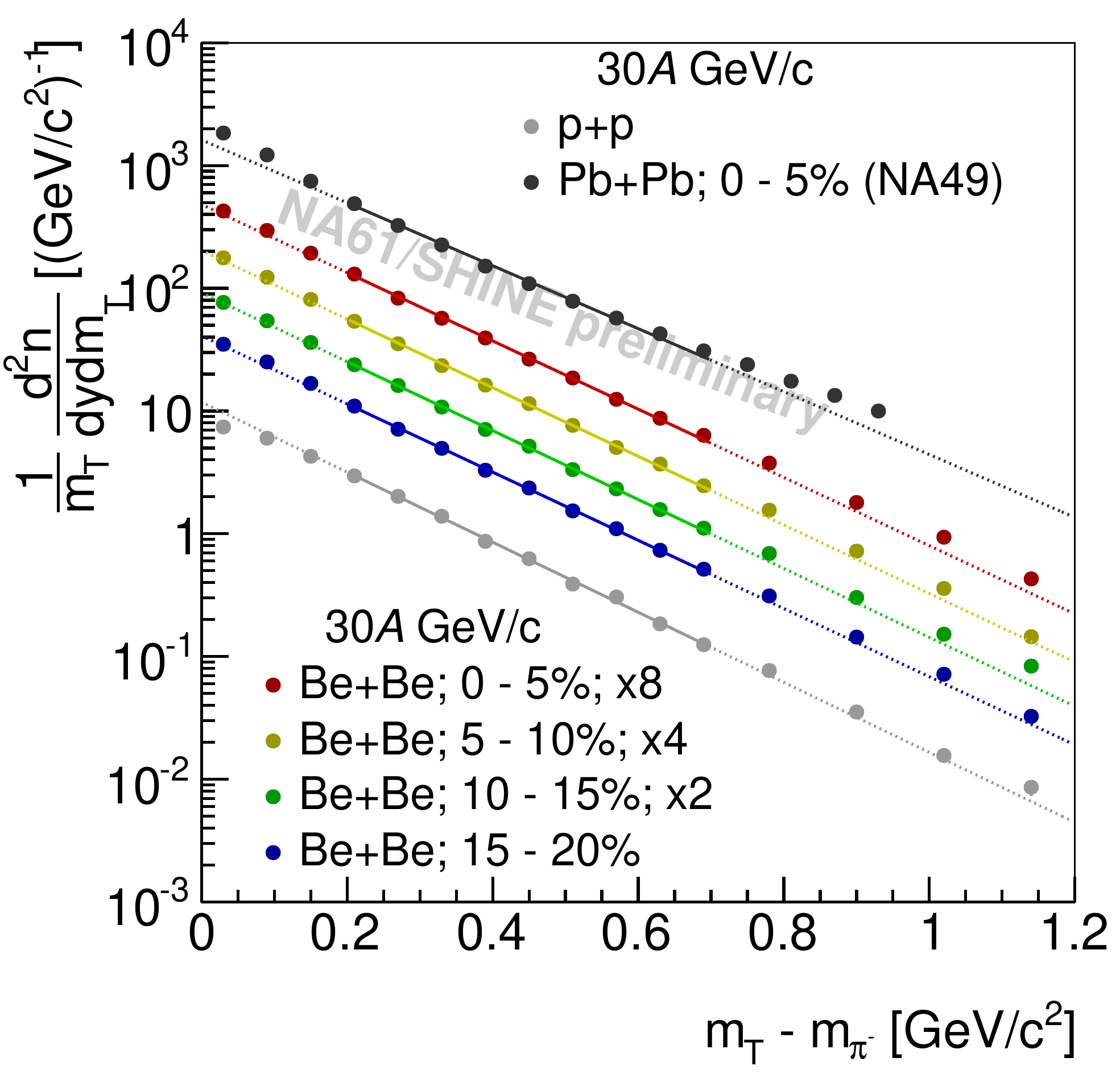}\end{minipage}
\begin{minipage}[t]{0.32\textwidth}\includegraphics[width=0.99\textwidth]{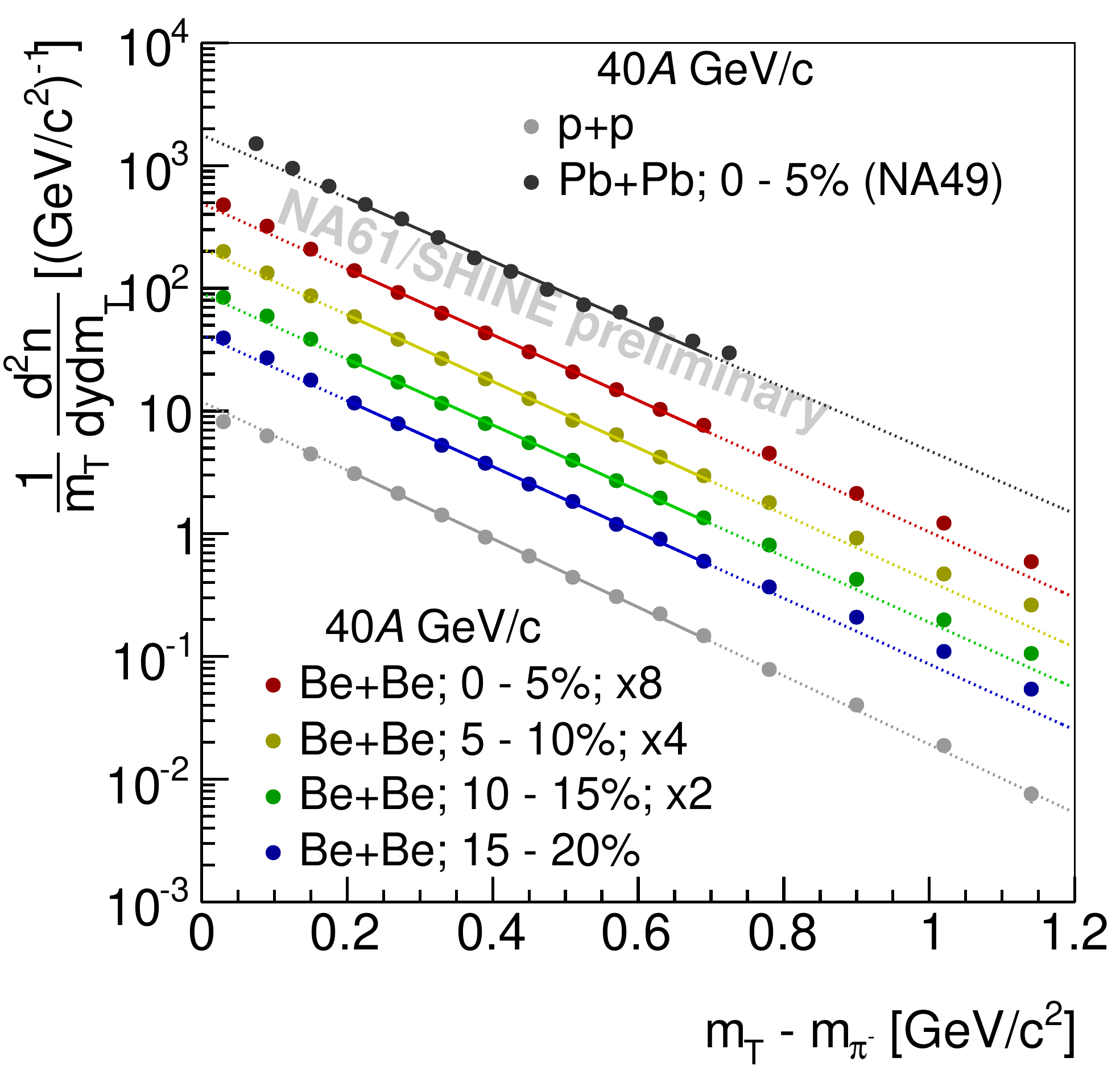}\end{minipage}
\begin{minipage}[t]{0.32\textwidth}\includegraphics[width=0.99\textwidth]{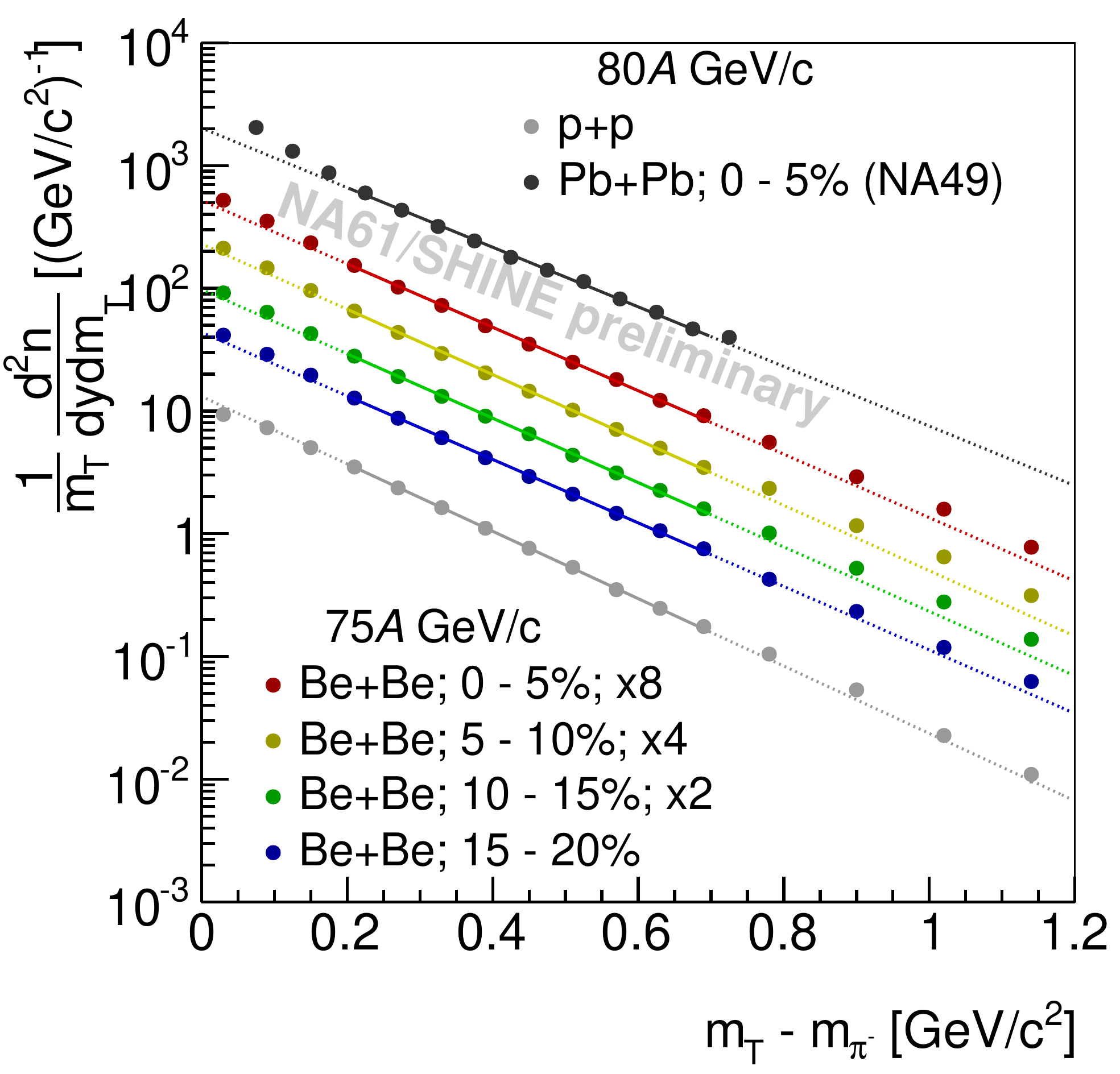}\end{minipage}
\begin{minipage}[t]{0.32\textwidth}\includegraphics[width=0.99\textwidth]{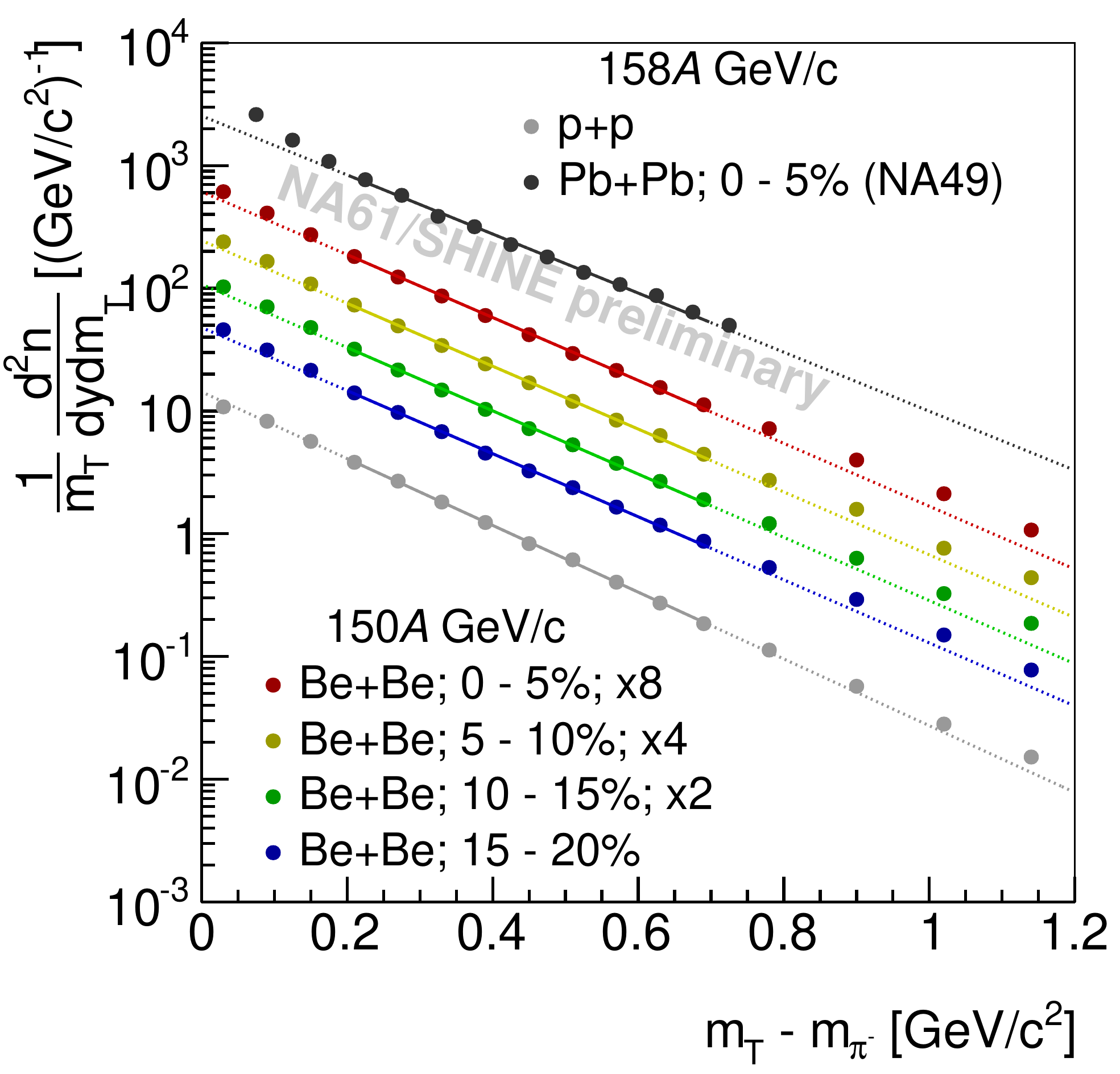}\end{minipage}
\end{center}
\caption{Transverse mass spectra of $\pi^-$ at mid-rapidity in $^7$Be+$^9$Be collisions at 5 beam momenta 20$A$, 30$A$, 40$A$, 75$A$, and 150$A$ \GeVc 
for 4 centrality classes compared
to p+p \cite{PionPaper} and central Pb+Pb \cite{LeadPaper} data at the nearest measured energy.}
\label{fig:mTSpectra} % Give a unique lebel
\end{figure}

Figure~\ref{fig:mTSpectra} presents the mid-rapidity $m_T$ spectra of $\pi^-$ production obtained in the 
analysis of $^7$Be+$^9$Be collisions at five beam momenta and for four centrality classes 
and compares them to data for inelastic p+p interactions as well as to Pb+Pb results of NA49.
The spectra are exponential for p+p reactions but start deviating from this simple shape for
collisions of nuclei.

\begin{figure}
\begin{center}
\resizebox{0.45\textwidth}{!}{
  \includegraphics{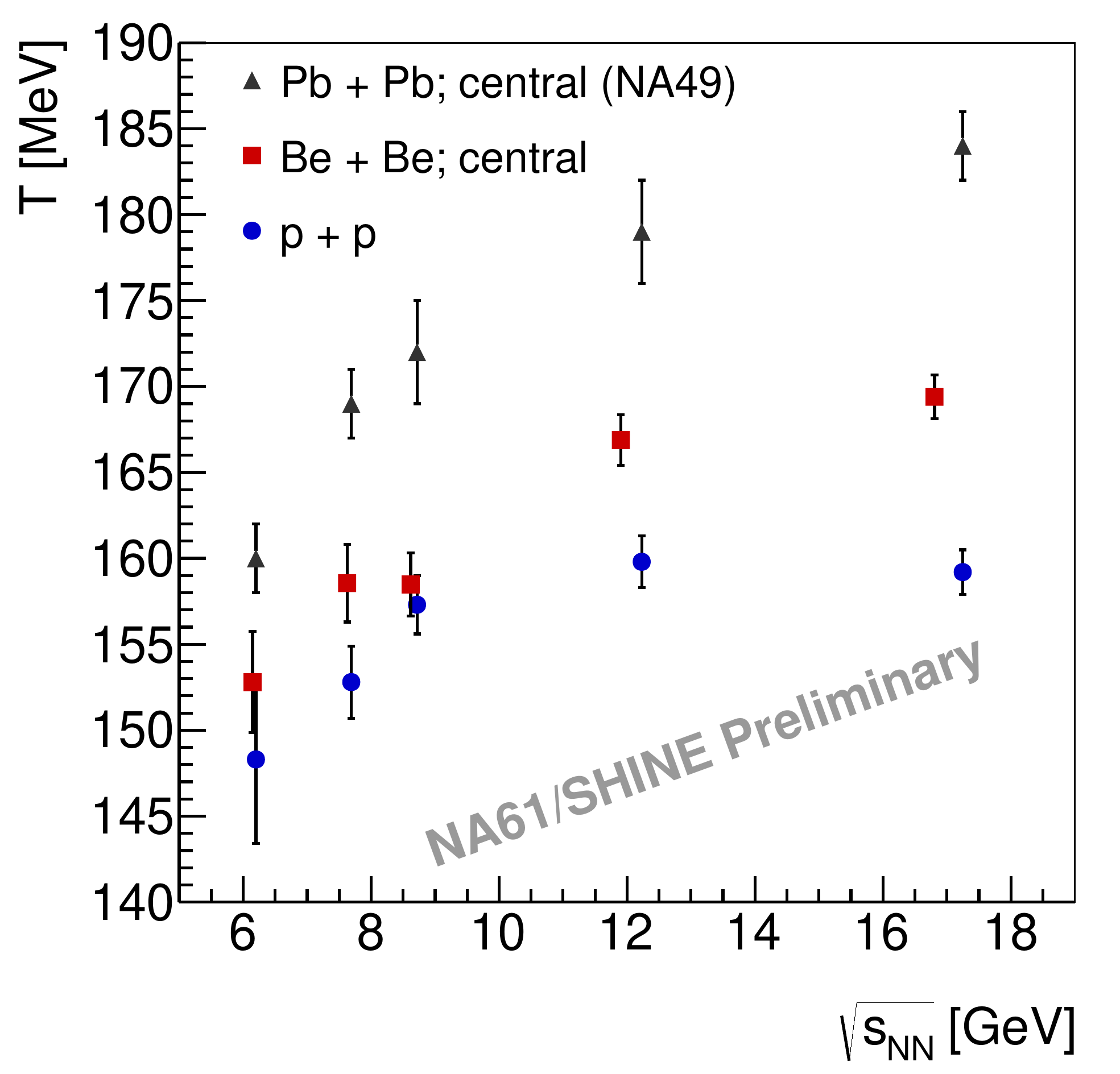}
}
\end{center}
\caption{Inverse slope parameter $T$ obtained from $\pi^-$ transverse mass distributions at mid-rapidity
for p+p \cite{PionPaper} and  central $^7$Be+$^9$Be and Pb+Pb \cite{LeadPaper} collisions.
The high inverse slope parameter in Pb+Pb interactions is due to radial flow. 
The values of $T$ for $^7$Be+$^9$Be are between those of p+p and Pb+Pb collisions and
can be interpreted as possible evidence of transverse collective flow in central $^7$Be+$^9$Be interactions.
Note: the fitted inverse slope parameter $T$ in A+A collisions is sensitive to the fit range and the location of the rapidity bin.
}
\label{fig:collectivity} % Give a unique lebel
\end{figure}

The inverse slope parameter $T$, characterising the spectra, was obtained from a fit to the data using the thermal ansatz:
\begin{equation}
\frac{d^2n}{m_Tdm_Tdy}=Ae^{-\frac{m_T}{T}}~.
\end{equation}
The dependence of $T$ on the collision energy (see Fig.~\ref{fig:collectivity}) for the most central $^7$Be+$^9$Be events was compared with the~\NASixtyOne data
on p+p interactions, as well as with the NA49 data on central Pb+Pb collisions. The results for $^7$Be+$^9$Be lie above
those for p+p and below those for Pb+Pb collisions leading to the conclusion that a collective
flow effect already starts in central $^7$Be+$^9$Be reactions. 

\subsection{The quest for the critical point}
When the freeze-out of the produced particle system occurs close to the critical point of strongly interacting matter, 
one should observe a maximum of event-by-event fluctuations. In order to compare the data for different
reactions and detector acceptances, the studied observables should be independent of 
system size and its fluctuations. Therefore, \NASixtyOne measured two strongly intensive measures $\Delta$ and $\Sigma$ 
proposed in Refs.~\cite{StronglyIntensive,StronglyIntensive2}.

\begin{figure}
\begin{center}
\resizebox{0.9\textwidth}{!}{
  \includegraphics{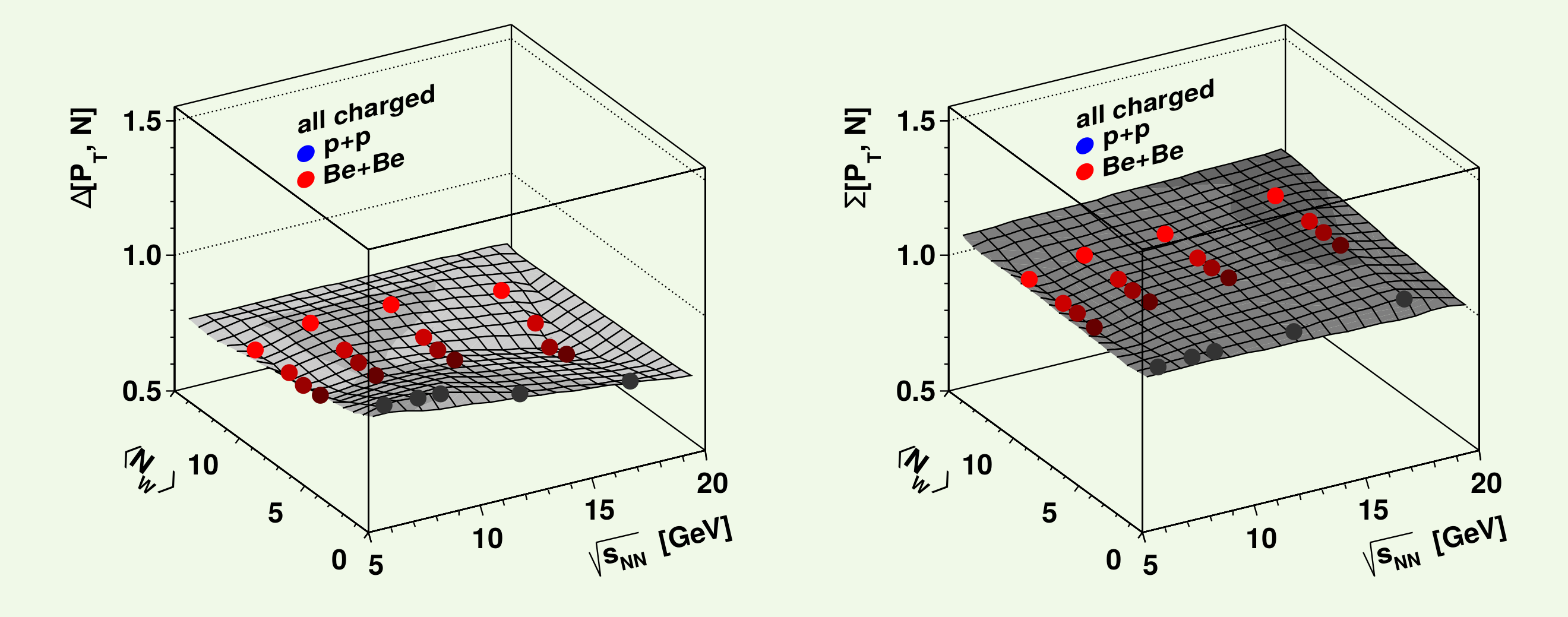}
}
\end{center}
\caption{The critical point of strongly interacting matter is expected to cause
a maximum of fluctuations of hadronic observables. No sign of such an anomaly is 
observed for $\Delta$ and $\Sigma$ in either p+p or $^7$Be+$^9$Be collisions.
The plotted measures $\Delta$ and $\Sigma$ of transverse momentum fluctuations are strongly
intensive, thus independent of the system volume and its fluctuations. }
\label{fig:fluctuations} % Give a unique lebel
\end{figure}

Preliminary results for charged particles (mainly pions) in four centrality classes of $^7$Be+$^9$Be reactions
are shown in Fig.~\ref{fig:fluctuations} and  compared to p+p data.
Consistent with expectations for such small-size systems,
no structure is observed in this energy - system size scan that might suggest effects of the critical point.

\section{Summary}
The \NASixtyOne experiment at the CERN SPS is a unique facility which operates with various primary and secondary beams (hadrons, ions)
interacting with stationary targets.

The scientific program of the experiment covers three main fields of interest: 
reference measurements for neutrino physics and cosmic ray experiments, as well as studies of hadron interactions
at beam momenta needed to reach the onset of deconfinement and the critical point of strongly interacting matter. 
In addition, cold nuclear matter effects are investigated in proton-nucleus interactions.

The ongoing \NASixtyOne scan program covers a wide range of beam energies and collision system sizes. First
results from p+p and $^7$Be+$^9$Be collisions  are presented in this paper. 

High precision double-differential pion spectra were measured at five different energies for both systems. 
In p+p reactions the energy dependence of the $K^+/\pi^+$ and $K^-/\pi^-$ ratios and 
of the inverse slope parameter $T$ of the transverse mass spectra of kaons were determined. The results for $K^+$ show 
rapid changes with collision energy, even for inelastic p+p interactions. These structures resemble similar effects 
observed in central Pb+Pb interactions at SPS energies. 

Collective flow effects are observed already in $^7$Be+$^9$Be collisions.

No sign for the critical point of strongly interacting matter was found for these small-sized systems.

\section*{Acknowledgements}
This work was supported by
the Hungarian Scientific Research Fund (grants OTKA 68506 and 71989),
the J\'anos Bolyai Research Scholarship of
the Hungarian Academy of Sciences,
the Polish Ministry of Science and Higher Education (grants 667\slash N-CERN\slash2010\slash0, NN\,202\,48\,4339 and NN\,202\,23\,1837),
the Polish National Science Centre (grants~\textbf{2011\slash03\slash N\slash ST2\slash03691}, 2012\slash04\slash M\slash ST2\slash00816 and 
2013\slash11\slash N\slash ST2\slash03879),
the Foundation for Polish Science --- MPD program, co-financed by the European Union within the European Regional Development Fund,
the Federal Agency of Education of the Ministry of Education and Science of the
Russian Federation (SPbSU research grant 11.38.193.2014),
the Russian Academy of Science and the Russian Foundation for Basic Research (grants 08-02-00018, 09-02-00664 and 12-02-91503-CERN),
the Ministry of Education, Culture, Sports, Science and Tech\-no\-lo\-gy, Japan, Grant-in-Aid for Sci\-en\-ti\-fic Research (grants 18071005, 19034011, 19740162, 20740160 and 20039012),
the German Research Foundation (grant GA\,1480/2-2),
the EU-funded Marie Curie Outgoing Fellowship,
Grant PIOF-GA-2013-624803,
the Bulgarian Nuclear Regulatory Agency and the Joint Institute for
Nuclear Research, Dubna (bilateral contract No. 4418-1-15\slash 17),
Ministry of Education and Science of the Republic of Serbia (grant OI171002),
Swiss Nationalfonds Foundation (grant 200020\-117913/1)
and ETH Research Grant TH-01\,07-3.
Finally, it is a pleasure to thank the European
Organisation for Nuclear Research for strong support and
hospitality and, in particular, the operating crews of the CERN
SPS accelerator and beam lines who made the measurements
possible.

\end{document}